\documentclass{emulateapj}
\usepackage{graphicx}
\usepackage{natbib}
\shorttitle{Peculiar Type Ic SN\,2005ek}
\shortauthors{M. R. Drout, et al.}

\begin{document}

\title{The Fast and Furious Decay of the Peculiar Type Ic Supernova 2005ek}

\author{M. R. Drout\altaffilmark{1}, A. M. Soderberg\altaffilmark{1}, P. A. Mazzali\altaffilmark{2,3,4}, J. T. Parrent\altaffilmark{5,6}, R. Margutti\altaffilmark{1}, D. Milisavljevic\altaffilmark{1}, N. E. Sanders\altaffilmark{1}, R. Chornock\altaffilmark{1}, R. J. Foley\altaffilmark{1}, R. P. Kirshner\altaffilmark{1}, A. V. Filippenko\altaffilmark{7}, W. Li\altaffilmark{7,8}, P. J. Brown\altaffilmark{9}, S. B. Cenko\altaffilmark{7}, S. Chakraborti\altaffilmark{1}, P. Challis\altaffilmark{1}, A. Friedman\altaffilmark{1,10}, M. Ganeshalingam\altaffilmark{7}, M. Hicken\altaffilmark{1}, C. Jensen\altaffilmark{1}, M. Modjaz\altaffilmark{11}, H. B. Perets\altaffilmark{12}, J. M. Silverman\altaffilmark{13,14}, D. S. Wong\altaffilmark{15}}

\altaffiltext{1}{Harvard-Smithsonian Center for Astrophysics, 60 Garden Street, Cambridge, MA 02138}
\altaffiltext{2}{Astrophysics Research Institute, Liverpool John Moores University, CH41 1LD Liverpool, UK}
\altaffiltext{3}{INAF-Osservatorio Astronomico di Padova, Vicolo dell'Osservatorio 5, I-35122 Padova, Italy}
\altaffiltext{4}{Max-Planck-Institut for Astrophysik, Karl-Schwarzschildstr. 1, D-85748 Garching, Germany}
\altaffiltext{5}{Department of Physics \& Astronomy, Dartmouth College, 6127 Wilder Lab, Hanover, NH 03755, USA}
\altaffiltext{6}{Las Cumbres Observatory Global Telescope Network, Goleta, CA 93117, USA}
\altaffiltext{7}{Department of Astronomy, University of California, Berkeley, CA 94720-3411, USA}
\altaffiltext{8}{Deceased 2011 December 12}
\altaffiltext{9}{Texas A\&M University, Department of Physics and Astronomy, College Station, TX 77843-4242, USA}
\altaffiltext{10}{Massachusetts Institute of Technology, Center for Theoretical Physics, 77 Massachusetts Ave., 6-304, Cambridge, MA 02139, USA}
\altaffiltext{11}{New York University, Center for Cosmology and Particle Physics, Department of Physics, 4 Washington Place, New York, NY 10003, USA}
\altaffiltext{12}{Physics Department, Technion - Israel Institute of Technology, Haifa, Israel 32000}
\altaffiltext{13}{Department of Astronomy, University of Texas at Austin, Austin, TX 78712, USA}
\altaffiltext{14}{NSF Astronomy and Astrophysics Postdoctoral Fellow}
\altaffiltext{15}{University of Alberta, Physics Dept, 4-183 CCIS, Edmonton, AB, T6G 2E1, CANADA}

\begin{abstract}
We present extensive multi-wavelength observations of the extremely rapidly declining Type Ic supernova, SN\,2005ek.  Reaching a peak magnitude of $M_R = -17.3$ and decaying by $\sim 3$ mag in the first 15 days post-maximum, SN\,2005ek is among the fastest Type I supernovae observed to date.  The spectra of SN\,2005ek closely resemble those of normal SN~Ic, but with an accelerated evolution.  There is evidence for the onset of nebular features at only nine days post-maximum.  Spectroscopic modeling reveals an ejecta mass of $\sim 0.3$ M$_\odot$ that is dominated by oxygen ($\sim 80$\%), while the pseudo-bolometric light curve is consistent with an explosion powered by $\sim 0.03$ M$_\odot$ of radioactive $^{56}$Ni.  Although previous rapidly evolving events (e.g., SN\,1885A, SN\,1939B, SN\,2002bj, SN\,2010X) were hypothesized to be produced by the detonation of a helium shell on a white dwarf, oxygen-dominated ejecta are difficult to reconcile with this proposed mechanism.  We find that the properties of SN\,2005ek are consistent with either the edge-lit double detonation of a low-mass white dwarf or the iron-core collapse of a massive star, stripped by binary interaction.  However, if we assume that the strong spectroscopic similarity of SN\,2005ek to other SN~Ic is an indication of a similar progenitor channel, then a white-dwarf progenitor becomes very improbable.  SN\,2005ek may be one of the lowest mass stripped-envelope core-collapse explosions ever observed.  We find that the rate of such rapidly declining Type I events is at least 1--3\% of the normal SN~Ia rate.

\end{abstract}

\email{mdrout@cfa.harvard.edu}

\section{Introduction}

The advent of dedicated supernova (SN) searches has dramatically increased the rate at which unusual transients are discovered.  In particular, high-cadence surveys have uncovered a diverse set of rapidly evolving events which reach SN luminosities (absolute magnitude between $-20$ and $-15$) but have observed properties that challenge the parameter space easily explained by traditional SN models (e.g., the collapse of the core of a massive star, or the thermonuclear disruption of a white dwarf).

The plethora of objects that have been referred to as ``rapidly evolving'' include both Type I (hydrogen poor) and Type II (hydrogen rich) events (see \citealt{Filippenko1997} for a review of traditional SN classifications). Although the main physical process leading to optical emission varies among supernovae, in all cases the characteristic timescale offers insight into the amount of participating material. Rapid evolution typically implies lower masses.  For supernovae powered by hydrogen recombination (e.g., Types IIP, IIL) and radioactive decay (e.g., Types Ia, Ib, Ic), rapid timescales indicate a low hydrogen envelope mass and a short photon diffusion timescale, respectively.  For supernovae powered by interaction with external gas (e.g., Type IIn), a rapid decline implies a steep decrease in circumstellar medium (CSM) density, and a short overall timescale implies a small radius over which this material is located.

\begin{figure}[!ht]
\includegraphics[width=\columnwidth]{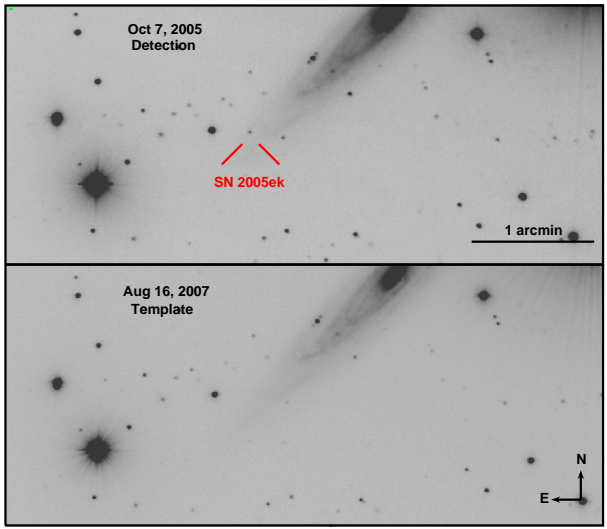}
\caption{\label{fig:host} \emph{Top:} R-band Palomar 60-inch (P60) image of SN\,2005ek, on the outskirts of UGC\,2526. The SN location is marked by red crosshairs. \emph{Bottom:} P60 template image of the region around SN\,2005ek, taken on 2007 Aug. 26. }
\end{figure}

Among the Type~I events labeled as rapidly evolving are the SN 1991bg-like SN~Ia \citep{Filippenko92,Leibundgut93}, the ``calcium-rich'' transients for which SN\,2005E is the prototype \citep{Perets2010,Kasliwal2012,Valenti2013}, and some members of the recently defined Type Iax supernovae \citep{Foley2013a}.  They earn the title ``rapidly evolving'' because they decay by 1--2 mag in the first 15 days post-maximum. Many of these objects possess peak luminosities lower than those of normal SN~I ($\gtrsim -15$ mag) and are thought to be powered by radioactive decay. Although their host galaxies are diverse, members of the first two classes above have exploded in elliptical galaxies. In addition, several  luminous (M $\lesssim -19$ mag) transients have been observed that decay on similar timescales, but show narrow hydrogen and/or helium emission lines in their spectra, indicating that they are at least partially powered by interaction with a dense CSM. These include the Type IIn SN PTF09uj \citep{Ofek2010} and the Type Ibn SN\,1999cq \citep{Matheson2000}.

However, the record for the most rapidly declining SN observed thus far does not belong to any of these objects.  SN\,2002bj \citep{Poznanski2010} and SN\,2010X \citep{Kasliwal2010} easily outstrip them, declining by $\gtrsim 3$ mag in the first 15 days post-maximum.  The ejecta masses inferred for these two events are very small ($\lesssim 0.3$ M$_\odot$), but their peak luminosities are within the typical range for SN~Ib/Ic ($-19 \lesssim$ M  $\lesssim -17$; \citealt{Drout2011}).  These two facts, coupled with the lack of hydrogen in their spectra (SN\,2002bj is a SN~Ib, SN\,2010X a SN~Ic), have led several authors to hypothesize that they were produced by the detonation of a helium shell on a white dwarf (a ``.Ia'' supernova; \citealt{Woosley1986,Chevalier1988,Bildsten2007,Shen2010,Waldman2011,Sim2012}).  

Two potential other members of this class include SN\,1885A and SN\,1939B \citep[see, e.g.,][]{Perets2011,Chevalier1988,deVaucouleurs1985,Leibundgut1991}. While both SN\,2002bj and SN\,2010X were found in star-forming galaxies, SN\,1939B exploded in an elliptical, a fact suggestive of an old progenitor system.  However, while the post-maximum decline rates of these objects are similar, even the well-studied events show differences in their other properties.  SN\,2002bj was $\sim 1.5$ mag brighter, significantly bluer, and exhibited lower expansion velocities than SN\,2010X. It has not yet been established whether all (or any) of these extremely rapidly declining objects belong to the same class of events.

Here we present the discovery and panchromatic follow-up observations of SN\,2005ek, another very rapidly declining and hydrogen-free event that closely resembles SN\,2010X.  In \S \ref{sec:Obs} we present our extensive multi-wavelength observations, while in \S \ref{sec:lc}, \S \ref{sec:SpecModel}, \S \ref{sec:explosion}, and \S \ref{sec:HostProps} we respectively describe the photometric and spectroscopic properties, explosion parameters, and host-galaxy environment of SN\,2005ek. Section~\ref{sec:rates} examines the rates of such transients. Finally, in \S \ref{sec:theories}, we discuss progenitor channels that could lead to such a rapidly evolving explosion.

\section{Observations}\label{sec:Obs}

\subsection{Discovery}

SN\,2005ek was discovered by the Lick Observatory Supernova Search (LOSS)  using the Katzman Automatic Imaging Telescope \citep[KAIT;][]{Filippenko01} on 2005 Sep.\ 24.53 (UT dates are used throughout this paper) with an unfiltered (clear) $m \approx 17.5$ mag.  The object was not detected in previous KAIT images on Sep.\ 18.51 to a limit of $m \approx 19$ mag, while subsequent imaging on Sep.\ 25.37 revealed that the transient had brightened to $m \approx 17.3$ mag \citep{i8604}.  SN\,2005ek is located in the outskirts of its host galaxy, UGC 2526, with distance $D =  66.6 \pm 4.7$ Mpc\footnote{We adopt the NED distance after correction for Virgo, Great Attractor, and Shapley Supercluster Infall and assuming $\rm H_0 = 73$ km s$^{-1}$ Mpc$^{-1}$ \citep{mhf+00}.} and morphology Sb.

\begin{figure*}[!ht]
\begin{centering}
\includegraphics[width=0.9\textwidth]{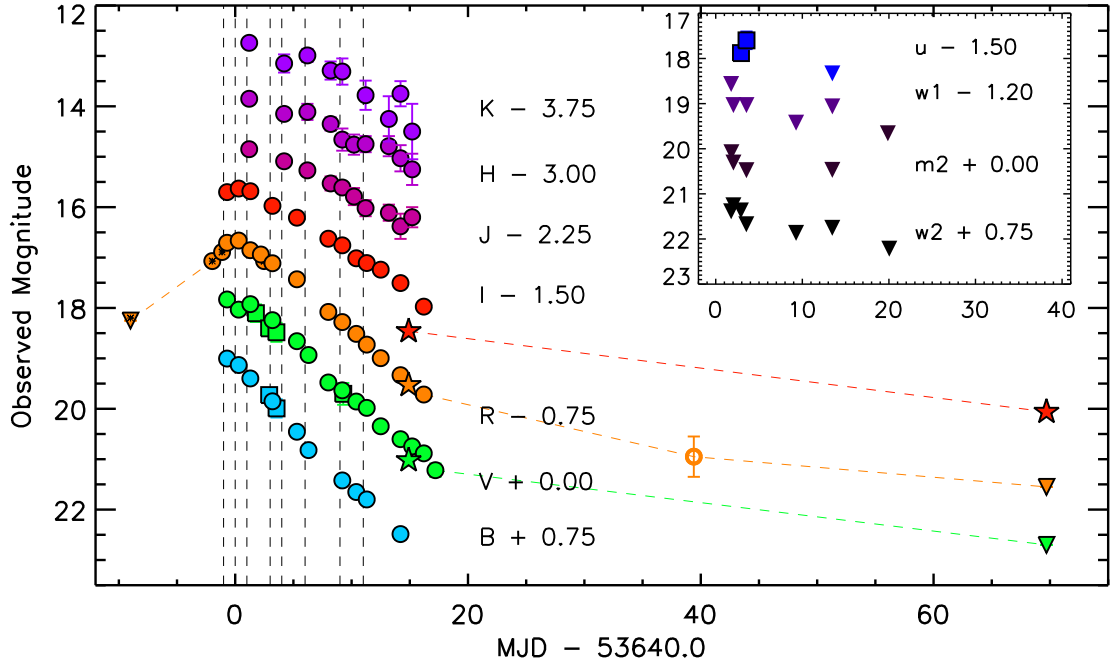}
\caption{\label{fig:Photom} Photometry of SN\,2005ek. P60 $BVRI$- and PAIRITEL $JHK$-band data (circles) are respectively shown as cyan, green, orange, red, and three shades of violet. Lick Observatory discovery observations and the pre-discovery nondetection are shown as orange circles and a triangle with an asterisk inside. P200 $gri$-band data are shown as green, orange, and red stars; triangles denote upper limits.  A late-time FLWO $r$-band detection is shown as an open orange circle. UVOT $b$- and $v$-band detections are shown as blue and green squares in the main panel.  UVOT $u$-, $w1$-, $m2$-, and $w2$ detections (squares) and upper limits (triangles) are shown in the inset. Vertical lines in the main panel indicate epochs on which spectroscopy was obtained.}
\end{centering}
\end{figure*}

\citet{wps+05} obtained a spectrum of SN\,2005ek on Sep.\ 26 with the Shane 3-m reflector (plus Kast spectrograph) at Lick Observatory and reported that SN\,2005ek was a ``young supernova, probably of Type Ic.'' After this spectroscopic identification, we promptly initiated a panchromatic follow-up program spanning the radio, infrared (IR), optical, ultraviolet (UV), and X-ray bands.

\subsection{Palomar 60-inch Imaging}\label{sec:p60}

We obtained nightly multi-band images of SN\,2005ek with the robotic Palomar 60-inch telescope (P60; \citealt{Cenko2006}) beginning on Sep.\ 26.3 and spanning through Oct.\ 15.2.  Each epoch consisted of 4--10 120~s frames in filters $B$, $V$, $R$, and $I$.  All P60 images were reduced with IRAF\footnote{IRAF is distributed by the National Optical Astronomy Observatory, which is operated by the Association for Research in Astronomy, Inc.\, under cooperative agreement with the National Science Foundation.} using a custom real-time reduction pipeline \citep{Cenko2006}.  Nightly images were combined using standard IRAF tools.   Images of the transient and host galaxy constructed from P60 data are shown in Figure~\ref{fig:host}.

For the P60 $V$ and $R$ bands we obtained template images of the region surrounding SN\,2005ek on 2007 Aug. 16 (bottom panel of Fig.~\ref{fig:host}), after the SN had faded from view.  We subtracted the host-galaxy emission present in the template image using a common point-spread function (PSF) method and then performed aperture photometry on the resulting difference images. For the $B$ and $I$ bands, no suitable template images were obtained, and we therefore performed PSF photometry on our stacked images directly.  A comparison of these two methods with our $V$- and $R$-band data revealed that the resulting photometry was consistent within the measured uncertainties.

In all cases, we measured the relative magnitude of the SN with respect to five field stars within the full $13' \times 13'$ P60 field of view.  Absolute calibration was performed based on Sloan Digital Sky Survey (SDSS) photometry of the field stars \citep{Ahn2012}, converted to the $BVRI$ system using the relations from \citet{Smith2002}.  Our resulting P60 photometry is listed in Table~\ref{tab:PhotomP60} and shown as filled circles in the main panel of Figure~\ref{fig:Photom}.  These data reveal that the light curve reaches maximum at $m_R \approx 17.4$ mag only $\sim 9$ days after the KAIT nondetection and subsequently decays very rapidly in all bands.

\subsection{Palomar 200-inch Imaging}

On 2005 Oct.\ 11 and Dec.\ 5 we imaged SN\,2005ek with the Large Format Camera (LFC) mounted on the Palomar 200-inch (5~m) telescope in the $g'$, $r'$, and $i'$ bands (120 s expsoures).  Image processing and PSF photometry were performed using standard packages within IRAF.  We performed our absolute calibration using SDSS photometry of field stars, in the same manner as described above.  Our resulting photometry is listed in Table~\ref{tab:Photom} and supplements the P60 data in the main panel of Figure~\ref{fig:Photom}. On our final epoch, the transient was only detected in the $i'$ band at $\sim 21.5$ mag. 

\subsection{Lick Observatory Imaging}

The LOSS unfiltered images of SN\,2005ek were reanalyzed for this work in the manner described by \citet{Li2003}.  These include the discovery images (Sep.\ 24.5 and 25.4) as well as an additional detection on Sep.\ 29.5 (orange circles with asterisks in Fig.~\ref{fig:Photom}; also see Table~\ref{tab:Photom}).  These detections indicate that SN\,2005ek was discovered on the rise, while our P60 observations, beginning on Sep.\ 26.3, show a continued rise for at most one day before a very rapid decline.

\subsection{FLWO Imaging}

We supplement our photometry with the $JHK$ data presented by Modjaz (\citeyear{Modjaz2007}; violet circles in Fig.~\ref{fig:Photom}).  These data were obtained on the PAIRITEL telescope at the Fred Lawrence Whipple Observatory (FLWO) and are well sampled over the same time period as the P60 observations described above. In addition, \citet{Modjaz2007} obtained an $r'$-band detection with the FLWO 1.2~m telescope on 2005 Nov.\ 5 (open orange circle; Fig.~\ref{fig:Photom}). 

\subsection{Swift UVOT Imaging }\label{sec:uvot}

\emph{Swift}-UVOT \citep{Roming05} observations of SN\,2005ek were triggered beginning on 2005 Sep.\ 29. Seven epochs were obtained in the $uvw2$, $uvm2$, $uvw1$, $u$, $b$, and/or $v$ filters over a time period of 20 days. The data were analyzed following the prescriptions of \cite{Brown09} and photometry is based on the UVOT photometric system of \cite{Poole08} with the sensitivity corrections and revised UV zeropoints of \citet{Breeveld2011}.  All data are listed in Table~\ref{tab:PhotomUVOT}. The $uvw2$-, $uvm2$-, $uvw1$-, and $u$-band data are shown in the inset of Figure~\ref{tab:Photom}, while the $b$- and $v$-band data are plotted as squares in the main panel.  The flux from SN\,2005ek appears to fall off in the blue, with only upper limits obtained in the UV bands.  The spectral energy distribution (SED) of SN\,2005ek will be examined in more detail in \S \ref{PB_LC}.

\subsection{Optical Spectroscopy }\label{sec:05ekSpec}

\begin{figure*}[!ht]
\begin{centering}
\includegraphics[width=0.96\textwidth]{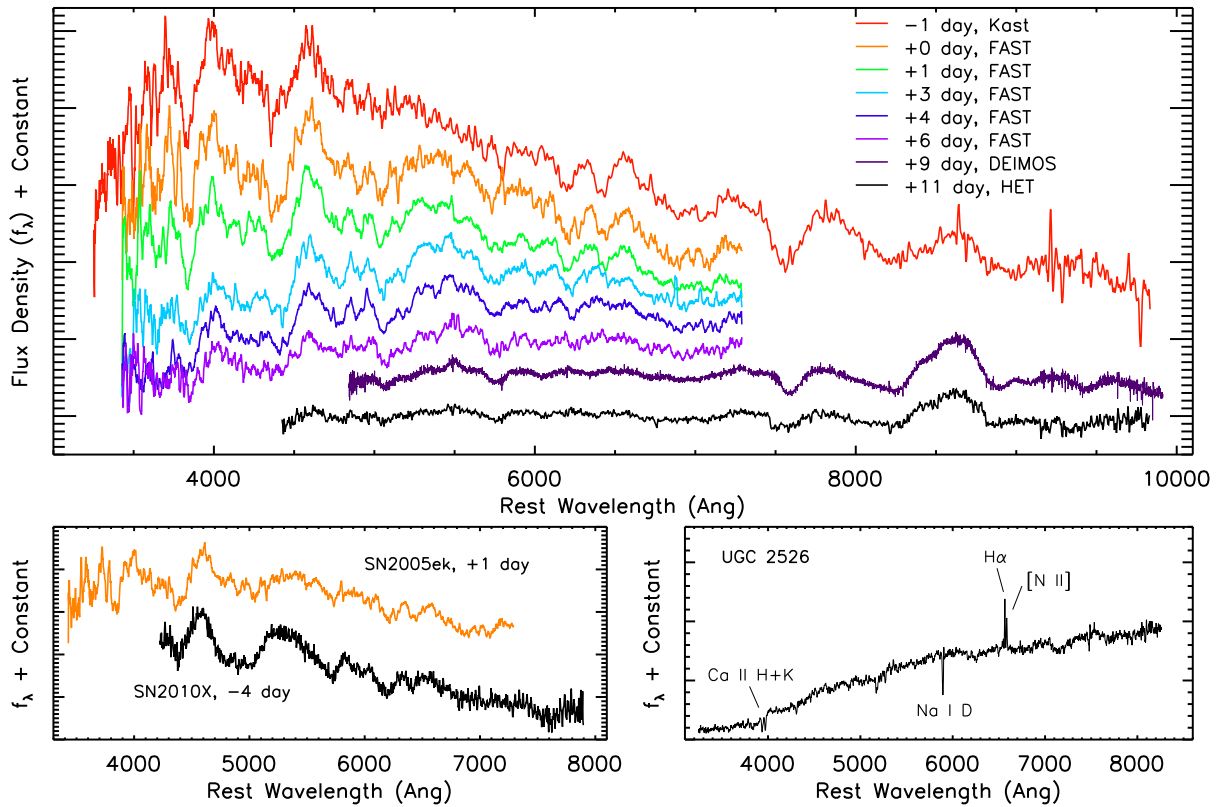}
\caption{\label{fig:SpecEvol} \emph{Top}: Optical spectra of SN\,2005ek (see text for details).  FAST and Kast spectra have been smoothed with windows of 11 \AA\ and 7 \AA, respectively. \emph{Lower left}: Comparison of the $+1$ day spectrum of SN\,2005ek with the $-4$ day spectrum of SN\,2010X \citep{Kasliwal2010}. \emph{Lower right:} Spectrum of the host galaxy, UGC\,2526. Major features are labeled.}
\end{centering}
\end{figure*}

Between 2005 Sep.\ 17 and 2005 Oct.\ 8, we obtained eight low-resolution spectra of SN\,2005ek from the FAST spectrograph \citep{Fabricant1998} on the FLWO 60-inch Tillinghast telescope, the Kast double spectrograph \citep{Miller93} on the 3-m Shane reflector at Lick Observatory, the DEIMOS spectrograph \citep{Faber03} mounted on the Keck-II 10-m telescope, and the 9.2-m Hobby-Eberly Telescope (HET) at McDonald Observatory. Technical details for all of our spectroscopic observations are summarized in Table~\ref{tab:Spectra}, and the epochs on which they were obtained are marked by dashed vertical lines in Figure~\ref{fig:Photom}.

The spectra were reduced in the manner described by \citet{Matheson2008}, \citet{Blondin2012}, and \citet{Silverman2012}.  Standard IRAF routines were used to subtract the overscan region and flatfield the two-dimensional CCD frames using a combined and normalized flatfield image. One-dimensional spectra were extracted and wavelength calibrated using comparison lamps obtained immediately following each exposure.  The FAST, Kast and DEIMOS spectra were flux calibrated utilizing a set of custom IDL routines which fit spectrophotometric standards to the data.  In addition, these routines apply a small shift to the wavelength calibration after cross-correlating night-sky lines with a template night-sky spectrum, apply a heliocentric correction, and use the spectrophotometric standards to remove telluric absorption features from the SN spectra (see, e.g., \citealt{Matheson2000}). The HET spectrum was flux calibrated in IRAF and no telluric correction was made.

All eight spectra are displayed in the top panel of Figure~\ref{fig:SpecEvol}, where the FAST and Kast spectra have been smoothed with windows of 11 \AA\ and 7 \AA, respectively.  A thorough analysis of the spectroscopic features and evolution of SN\,2005ek will be performed in \S \ref{sec:SpecModel}. The spectra closely resemble those of the rapidly declining SN\,2010X (bottom-left panel of Fig.~\ref{fig:SpecEvol}).

We obtained two spectra of the host galaxy, UGC 2526, with the Blue Channel Spectrograph \citep{Schmidt1989} on the 6.5-m MMT on 2011 Feb. 22 and 23.  The first observation was positioned on the explosion site, while the second was on the galactic nucleus and aligned along the major axis of the galaxy.  No strong nebular emission lines were detected in the explosion-site spectrum. Reduction, extraction, flux calibration, and telluric correction were performed in the manner described above, and the final spectrum centered on the galaxy nucleus is shown in the bottom-right panel of Figure~\ref{fig:SpecEvol}.  Weak, narrow H$\alpha$ and [N~II] emission lines are evident, along with a red continuum.

\subsection{Radio Observations }\label{sec:vla}

\begin{figure*}[!ht]
\begin{center}
\includegraphics[width=0.67\textwidth]{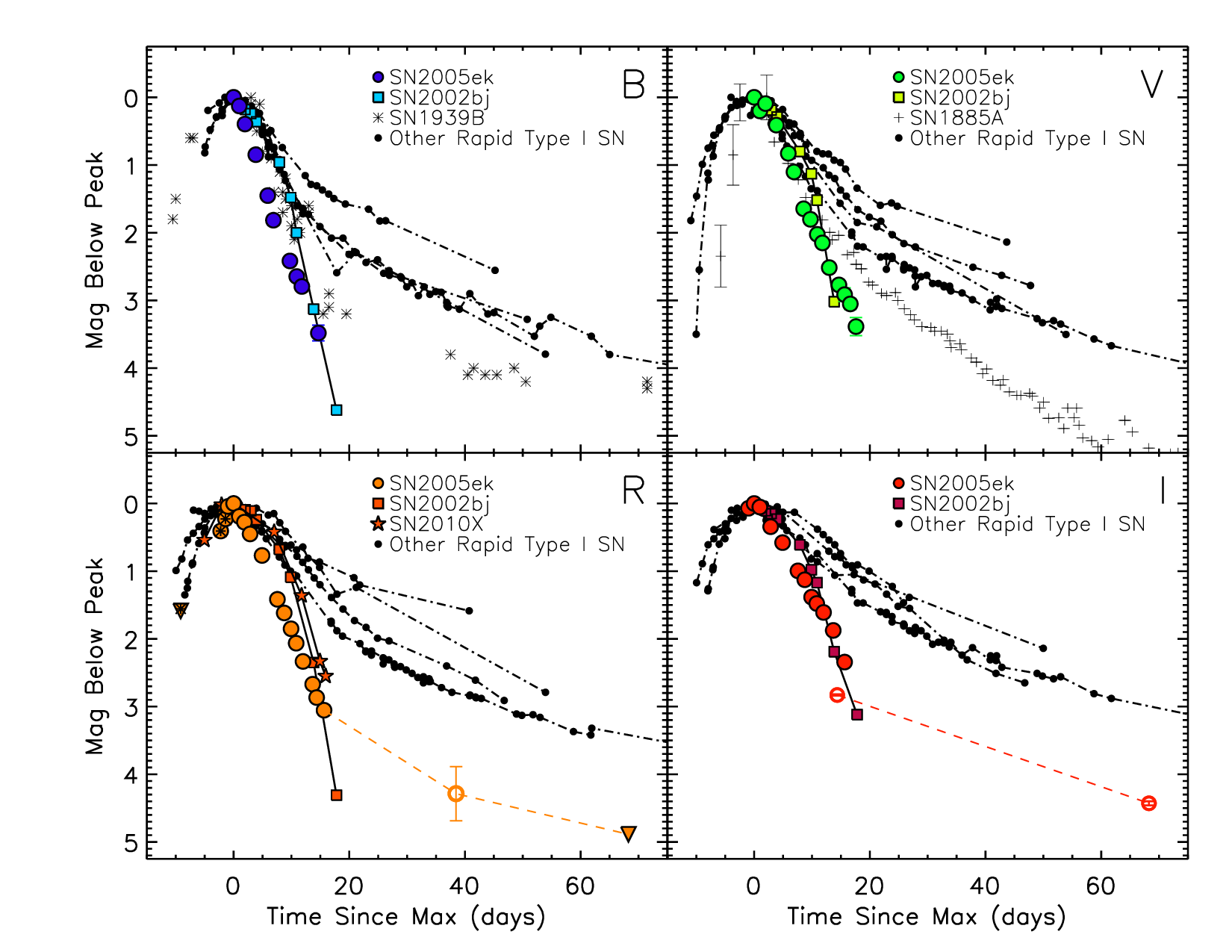}
\raisebox{1.2mm}{\includegraphics[width=0.3\textwidth]{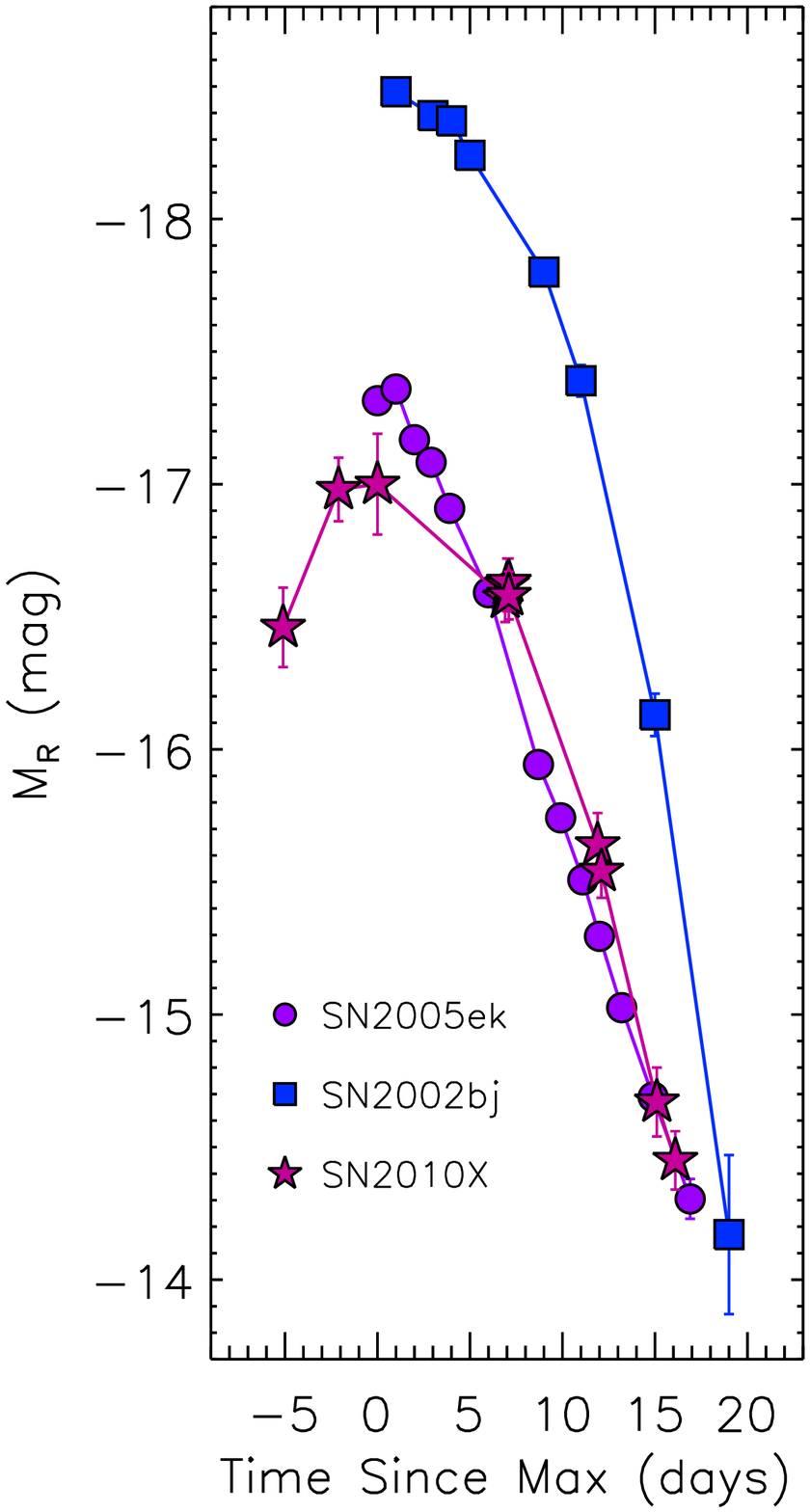}}
\caption{\label{fig:AppPan} \emph{Left:} Light curve of SN\,2005ek (colored circles), normalized to peak magnitude/epoch and compared to other rapidly evolving supernovae of Type I. The other events are SN\,1994I, SN\,2008ha, SN\,1998de, SN\,2005E (black circles), SN\,2002bj (colored squares), and SN\,2010X (colored stars; $R$-band panel only). \emph{Right:} $R$-band absolute magnitude light curves of SN\,2005ek (purple circles), SN\,2002bj (blue squares), and SN\,2010X (magenta stars). When error bars are not visible they are smaller than the plotted points.}
\end{center}
\end{figure*}

We observed SN\,2005ek with the Very Large Array (VLA) on 2005 Sep.\ 29 under our Target-of-Opportunity program to study the nonthermal properties of local Type Ib/Ic supernovae\footnote{VLA Intensive Survey of Naked Supernovae; VISiONS \citep{ams07}.} .  At 8.46 GHz we did not detect a coincident radio source, and we place an upper limit of $F_{\nu} \lesssim 128~\mu$Jy ($2\sigma$) on the flux density. At a distance of $\sim 67$ Mpc, this corresponds to a radio luminosity of $L_{\nu}\lesssim 7 \times 10^{26}~\rm ergs~s^{-1}~Hz^{-1}$.   This value is a factor of $\sim 7$ above the peak radio luminosity observed for the subluminous radio supernova SN\,2007gr \citep{Soderberg2010}.  

\subsection{X-ray Observations }\label{sec:X-ray}

SN\,2005ek was also observed with the X-ray Telescope \citep[XRT;][]{Burrows05}) onboard \emph{Swift} \citep{Gehrels04} beginning on 2005 Sep.\ 28.  The data were analyzed using the latest version of the HEASOFT package available at the time of writing (v. 6.13) and corresponding calibration files. Standard filtering and screening criteria were applied. All XRT data were coadded, resulting in a final 13~ks map spanning 19 days (median time of arrival = 7.44 days). No X-ray source is detected coincident with SN\,2005ek with a $3\sigma$ upper limit of $F_{x}<1.3\times 10^{13}~\rm{ergs\,s^{-1}\,cm^{-2}}$ (unabsorbed, 0.3--10 keV energy band). The Galactic neutral hydrogen column density in the direction of the SN is $1.0\times10^{21}~\rm{cm^{-2}}$ \citep{Kalberla05}. At $D \approx 67$ Mpc this yields a $3\sigma$ limit on the luminosity of  $L_{\nu}\lesssim 6.9 \times 10^{40}~\rm ergs~s^{-1}~Hz^{-1}$.  This value lies above the peak luminosity level for all but the most X-ray loud SN~I at a similar time after explosion (e.g., SN\,1998bw; \citealt{Pian2000}) .

\section{Light-Curve Properties }\label{sec:lc}

\subsection{Reddening }\label{dred}

Reddening due to the Milky Way in the direction of UGC 2526 has a value of $E(B-V) = 0.210$ mag, according to the infrared dust maps of \citet{Schlegel1998}. In order to estimate the host-galaxy contribution to the total reddening, we examine our spectra for evidence of narrow Na~I~D lines, which have been shown to correlate with extinction due to dust \citep{Turatto2003,Poznanski2012}. It is only in our highest signal-to-noise ratio (S/N) spectra (DEIMOS, +9 days) that we see weak Na~I~D absorption at the redshift of UGC 2526\footnote{The large Na~I~D absorption seen in Figure~\ref{fig:SpecEvol} is along the line of site to the galaxy core.} with an equivalent width EW$_{\mathrm{Na~I~D}} \sim 0.31$ \AA.  Using the empirical relation of \citet{Poznanski2012}, this implies $E(B-V)_{\mathrm{host}} \approx 0.03$ mag. Given the low level of this inferred effect, combined with the uncertainties in the Na~I~D relation (e.g., \citealt{Poznanski2011}), we incorporate this value into our error budget for $E(B-V)_{\mathrm{tot}}$.  Throughout this paper we adopt an $R_V = A_V/E(B-V) = 3.1$ Milky Way extinction curve with a total reddening value of $E(B-V) = 0.210^{+0.036}_{-0.006}$ mag.

\subsection{Optical Light-Curve Evolution} \label{lcs}

\begin{figure*}[!ht]
\begin{centering}
\includegraphics[width=0.9\textwidth]{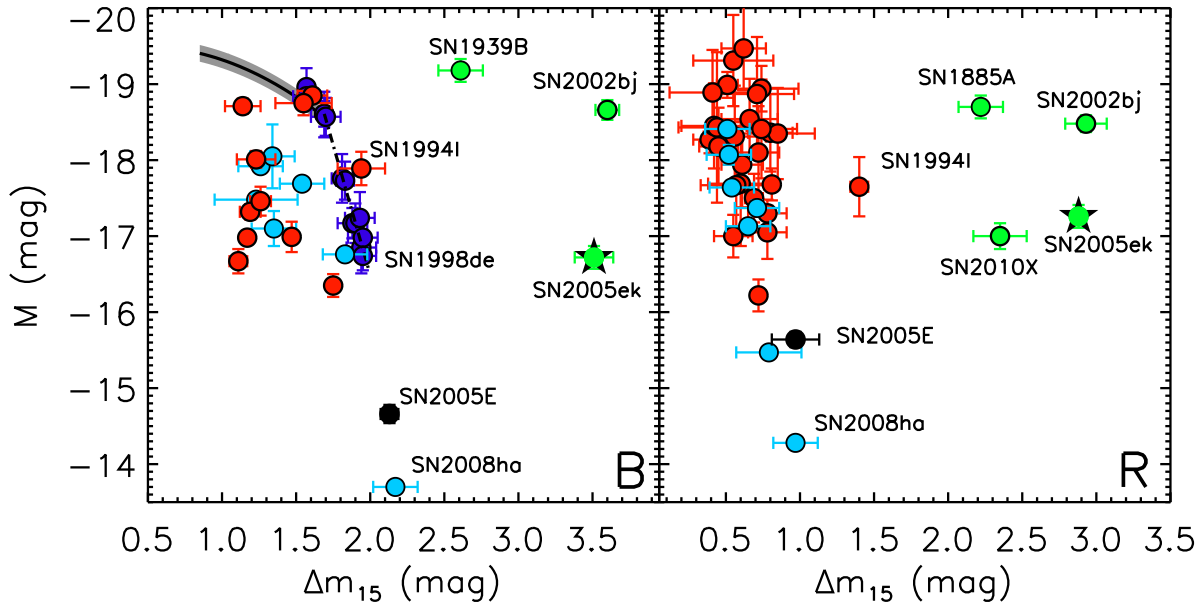}
\caption{\label{fig:DeltaM15} Absolute magnitude versus $\Delta m_{15}$ for a variety of SN in the $B$ band (left panel) and $R$ band (right panel). Normal SN~Ia are represented by a black line and shaded region.  SN\,1991bg-like SN~Ia are shown in dark blue, SN~Iax are shown in light blue, SN~Ib/Ic are shown in red, SN\,2005E is shown in black, and SN\,2002bj, SN\,2010X, SN\,1885A, SN\,1939B and SN\,2005ek are shown in green. SN\,2005ek is highlighted by a star.  Note the observations obtained for SN\,1885A are actually closer to modern day V-band. Objects plotted in Figure~\ref{fig:AppPan} are labeled.}
\end{centering}
\end{figure*}

In the left four panels of Figure~\ref{fig:AppPan} we display our $BVRI$ light curves, normalized to peak magnitude and epoch along with the $BVRI$ light curves for other SN~I which have previously been referred to as fast or rapidly evolving (black circles).  These include the rapid Type Ic SN\,1994I \citep{Richmond1996}, the ``calcium-rich'' Type Ib SN\,2005E \citep{Perets2010}, the SN\,1991bg-like Type Ia SN\,1998de \citep{Modjaz2001}, and the extremely low-luminosity SN\,2008ha \citep{Foley2009,Foley2010,Valenti2009}. Also shown are SN\,2010X (\citealt{Kasliwal2010}; colored stars), SN\,2002bj (\citealt{Poznanski2010}; colored squares), SN\,1885A (\citealt{deVaucouleurs1985,Perets2011}; plus signs), and SN\,1939B (\citealt{Leibundgut1991,Perets2011}; asterisks).

From Figure~\ref{fig:AppPan} it is clear that SN\,2005ek is an outlier even among rapid Type~I supernovae, decaying by $\gtrsim 3$ mag in 15 days and showing an unusually linear decline immediately post-maximum.  However, our final $r'$ and $i'$ detections do show evidence for a change in slope around 20--30 days post-maximum.  We can place an upper limit of $\sim 0.029$ mag day$^{-1}$ on the late-time $i'$-band slope of SN\,2005ek by comparing the two P200 $i'$ detections.  Both the timing of this transition and the late-time slope are comparable to those of the other rapid SN~I plotted in Figure~\ref{fig:AppPan}, although SN\,2005ek decays by 1--2 mag more before settling onto this late-time tail.

Basic properties for the $BVRIJHK$ bands are given in Table~\ref{tab:PhotomProps}.  We find the $R$-band peak epoch by fitting a low-order polynomial to the P60 $R$-band light curve supplemented with the Lick unfiltered photometry (which most closely mimics and is calibrated to the $R$ band; \citealt{Li2003}). This yields a peak epoch (MJD) of 53639.9 $\pm$ 0.3 day.  Unless otherwise noted, all phases throughout this paper are in reference to $R$-band maximum.  After correcting for distance and reddening we derive peak absolute $BVRI$ magnitudes ranging from $-16.72 \pm 0.15$ ($B$ band) to $-17.38 \pm 0.15$ ($I$ band).  This places SN\,2005ek at a peak optical magnitude very similar to SN\,2010X and $\sim 1.5$ mag below SN\,2002bj, SN\,1885A, and SN\,1939B.  In the right panel of Figure~\ref{fig:AppPan} we compare the absolute $R$-band light curves of SN\,2005ek, SN\,2010X, and SN\,2002bj.

In order to quantify the rapid decline of SN\,2005ek we calculate the time over which the magnitude declines by a factor of $e^{-1}$ ($\tau_e$), the number of magnitudes the light curve declines in the first 15 days past maximum ($\Delta m_{15}$), and the linear decline rate in magnitudes per day.  The first two quantities are calculated by linearly interpolating our data, and are measured with respect to the observed peak magnitude/date.  The decline rates are estimated from linear least-square fits to the data between +2 and +16 days.  Uncertainties for all properties listed in Table~\ref{tab:PhotomProps} were estimated using a Monte Carlo technique to produce and analyze 1000 realizations of our data\footnote{Each data point in each realization is a random variable chosen from a normal distribution with a mean and variance determined by its counterpart in our initial data set.}.

Our best constraint on the rise time of SN\,2005ek comes from the nondetection in a LOSS search image obtained on Sep.\ 18.5 (only $\sim 9$ days before the observed $R$-band maximum).  Despite this relatively short time frame, the upper limit of $\sim 19$ mag only moderately constrains the explosion epoch (Figure~\ref{fig:Photom}).  We can infer that SN\,2005ek rose slightly faster than its initial decline.

One of the most distinctive features of SN~Ia is the tight correlation between light-curve peak magnitude and decay rate.  This is in stark contrast to SN~Ib/Ic, which have been shown to fill a large portion of this parameter space \citep{Drout2011}.  In Figure~\ref{fig:DeltaM15} we plot peak absolute magnitude versus $\Delta m_{15}$ for SN\,2005ek and other supernovae of Type I.

\begin{figure}[!ht]
\includegraphics[width=\columnwidth]{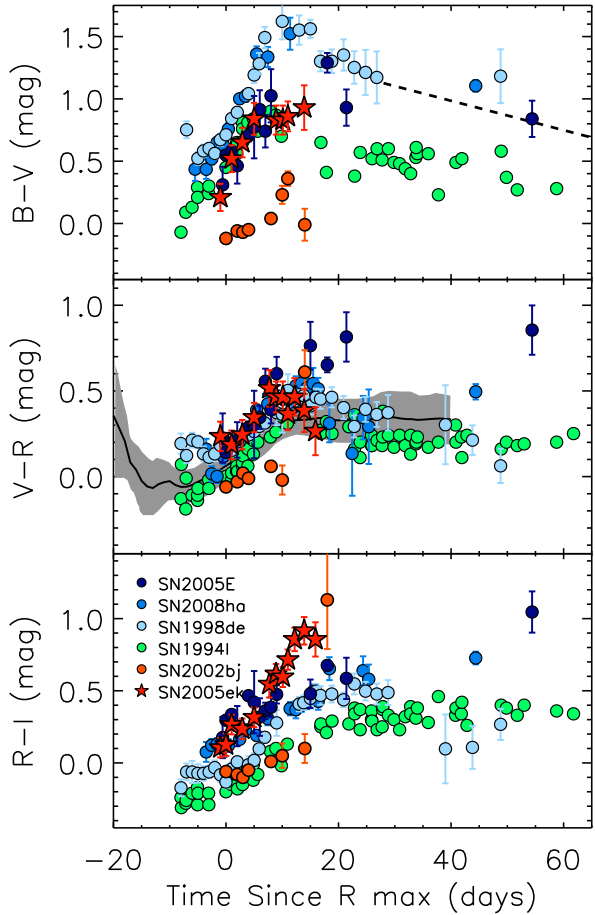}
\caption{\label{fig:Colors} $B-V$, $V-R$, and $R-I$ color evolution for SN\,2005ek (stars) and other rapidly evolving SN (circles).  See text for details.}
\end{figure}

The left panel of Figure~\ref{fig:DeltaM15} displays these values as measured in the $B$ band, which allows for a comparison with the well-studied \citet{Phillips93} relation for normal SN~Ia (\citealt{Phillips1999}; grey shaded region and solid black line) and the steeper relation found by \citet{Taubenberger2008} (dark blue points and dashed black line) for ``fast'' SN~Ia, as well as SN\,1939B.  Also shown in this panel are the literature sample of SN~Ib/Ic from \citet{Drout2011} (red points), the SN~Iax sample from \citet{Foley2013a} (light blue points), and several other peculiar Type~I events.  The right panel, measured in the $R$ band, allows for a comparison to the full sample of SN~Ib/Ic from \citet{Drout2011} as well as SN\,2010X (for which $r$ was the only well-sampled band obtained). We also include SN\,1885A which was observed in a photographic band which most closely resembles modern day V-band.  SN\,2005ek falls well outside the phase space covered by normal SN~Ib/Ic and is inconsistent with a simple extrapolation of either of the two SN~Ia scaling relations.

\subsection{Color Evolution }\label{Color}
In Figure~\ref{fig:Colors} we plot the $B-V$, $V-R$, and $R-I$ colors for SN\,2005ek, along with the color evolution for other rapidly evolving events. Also displayed is the Lira relation \citep{Phillips1999} which describes the remarkably similar $B-V$ color evolution for SN~Ia between 30 and 90 days past $V$-band maximum (dashed line; top panel).   In the middle panel we also show the SN~Ib/Ic template color curve from \citet{Drout2011} (grey shaded region).  \citet{Drout2011} demonstrated that dereddened SN~Ib/Ic show a very similar $V-R$ color evolution with a minimum dispersion at $\sim 10$ days post-maximum.

In the first $\sim 20$ days post-maximum, the $B-V$ and $V-R$ colors of SN\,2005ek appear consistent with those of other rapidly evolving SN~I: they exhibit a steady reddening with time and tentative evidence for the onset of a plateau between $\sim 10$ and 15 days.  Although the $R-I$ colors of SN\,2005ek also show a steady reddening, they do so at a much steeper rate than the other events displayed in Figure~\ref{fig:Colors}.  Because this unusual evolution is present only in $R-I$, it cannot be fully explained by a rapid cooling of the SN ejecta, but must be caused, in part, by a strong spectroscopic feature.  Indeed, from Figure~\ref{fig:SpecEvol} we see that by +9 days the emission component of the \ion{Ca}{2} near-IR triplet (which falls solidly inside the $I$ band) has grown substantially.  

Although the colors of SN\,2005ek are broadly consistent with those of other SN~I, they vary substantially from those of SN\,2002bj.  SN\,2002bj appears much bluer than any other object in $B-V$ and actually shows very little color evolution in either $V-R$ or $R-I$ until the final epoch, when it drastically reddens.  

\subsection{Spectral Energy Distribution}\label{SED}
\begin{figure}[!t]
\includegraphics[width=\columnwidth]{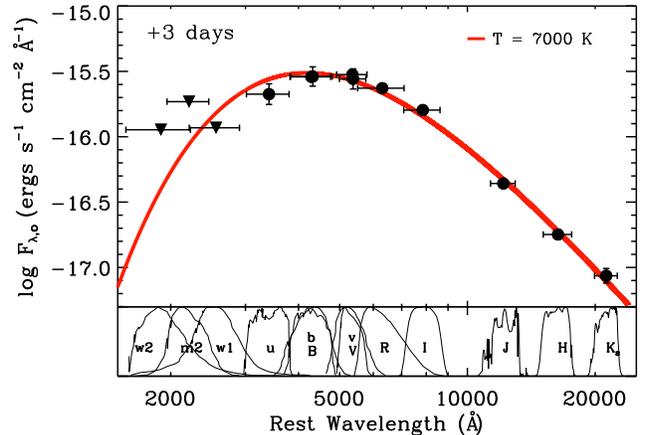}
\caption{\label{fig:SED} UV through IR spectral energy distribution of SN\,2005ek at three days post-maximum. Upper limits are indicated as triangles and bandpass shapes are shown in the lower panel.  The best-fitting 7000 K blackbody is shown as a red line.}
\end{figure}

In Figure~\ref{fig:SED} we plot the UV-optical-IR spectral energy distribution (SED) of SN\,2005ek from Sep.\ 30 (3 days post-maximum). Like many SN~I near maximum brightness, the SED of SN\,2005ek peaks in the optical.

Blackbody fits to various portions of the SED yield temperatures clustered around 7000 K (Fig.~\ref{fig:SED}, red curve).  This likely represents a lower limit on the true temperature due to the strong UV line blanketing produced by iron-peak elements in the spectra of SN~I (see, e.g., \citealt{Mazzali1993,Bongard2008}). Detailed modeling of the photospheric spectra reveals an ionization temperature of 9000--10,000 K near maximum brightness (\S \ref{sec:SpecModel}).

\subsection{Pseudo-Bolometric Light Curve}\label{PB_LC}

\begin{figure}[!ht]
\includegraphics[width=\columnwidth]{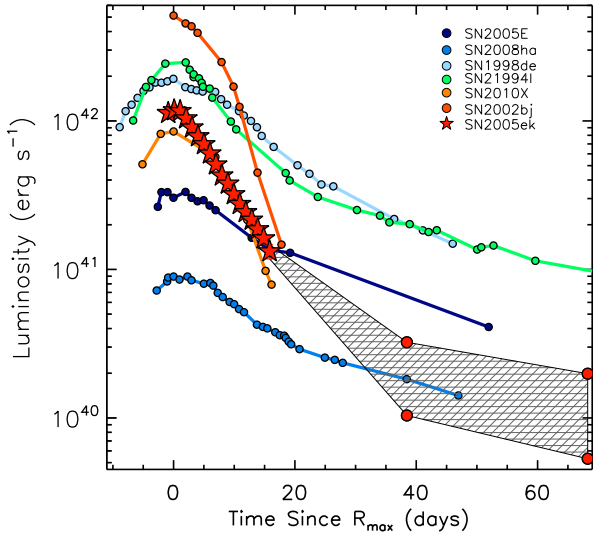}
\caption{\label{fig:PseudoB} Pseudo-bolometric light curve for SN\,2005ek (red stars) along with those of several other rapidly evolving supernovae (circles).  The red circles show our constraints on the late-time pseudo-bolometric luminosity of SN\,2005ek based on single-band detections.  The hatched region is meant to guide the eye.}
\end{figure}

To construct a pseudo-bolometric light curve we sum our observed $BVRIJHK$ data by means of a trapezoidal interpolation and attach a blackbody tail with a temperature and radius found by fitting a Planck function to the data. For epochs where we do not possess $JHK$ data we add a factor to our summed $BVRI$ data such that the total IR contribution ranges from $\sim 20$\% near maximum to $\sim 40$\% at the end of our observations \citep{Valenti2007}.  We do not include a UV correction as our {\it Swift} UV observations contain only upper limits.  Using this method, we find a peak bolometric luminosity of $(1.2 \pm 0.2) \times 10^{42}$ ergs s$^{-1}$ and a total radiated energy between $-1$ and +16 days of $(8.2 \pm 0.3) \times 10^{47}$ ergs.  

We also use the $r'$-band detection at +38 days and the $i'$-band detection at +68 days to place constraints on the late-time pseudo-bolometric evolution of SN\,2005ek.  First, we sum the observed flux at each epoch over the width of the appropriate filter to yield estimates for the minimum bolometric luminosity at +38 and +68 days. Second, by comparing the luminosity contained in our P200 $r'i'$-band observations at +15 days to our inferred bolometric luminosity at that epoch, we find that at +15 days the $r'$ and $i'$ bands contained $\sim 15$\% and $\sim 14$\% of the bolometric luminosity, respectively.  We use this fact to estimate a maximum pseudo-bolometric luminosity at +38 and +68 days under the assumption that the $r'$ and $i'$ contributions to the total luminosity will continue to increase.  The true bolometric luminosity likely lies closer to the higher of these two constraints.

In Figure~\ref{fig:PseudoB} we plot our pseudo-bolometric light curve of SN\,2005ek (red stars) along with the pseudo-bolometric light curves of SN\,1994I \citep{Richmond1996}, SN\,2008ha \citep{Moriya2010}, SN\,2002bj \citep{Poznanski2010}, SN\,2010X \citep{Kasliwal2010}, SN\,2005E, and SN\,1998de.  These last two were constructed in the manner described above from the photometry of \citet{Perets2010} and \citet{Modjaz2001}. The bolometric curve of SN\,2010X was constructed by computing $\nu F_{\nu}$ in the $r$ band.  A similar analysis of SN\,2005ek yields a pseudo-bolometric curve which is broadly consistent with our analysis above, but declines slightly more rapidly.  The hatched region in Figure~\ref{fig:PseudoB}  represents our constraints on the late-time pseudo-bolometric evolution of SN\,2005ek.

\section{Spectroscopic Properties }\label{sec:SpecModel}

\begin{figure}[!ht]
\includegraphics[width=\columnwidth]{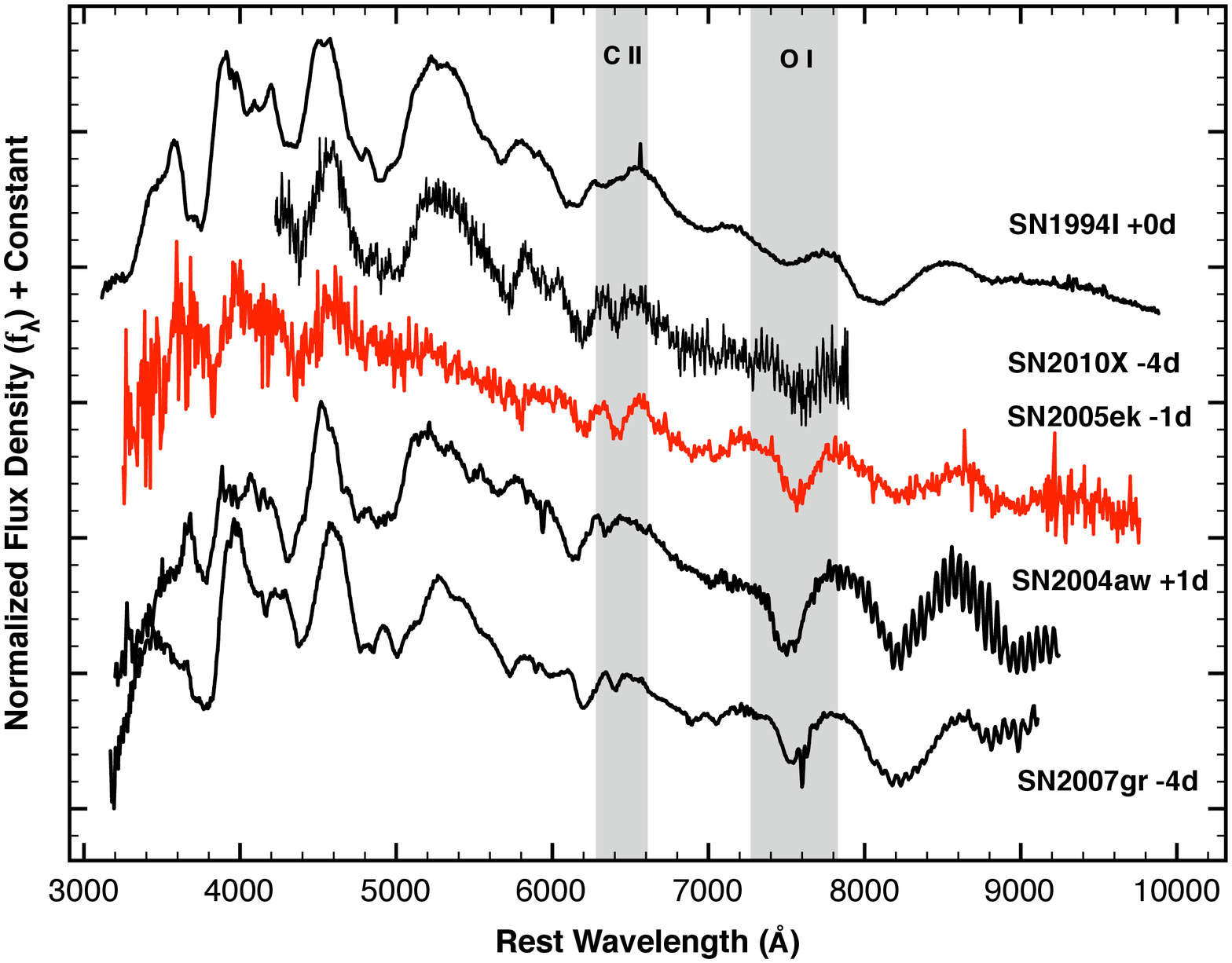}
\includegraphics[width=\columnwidth]{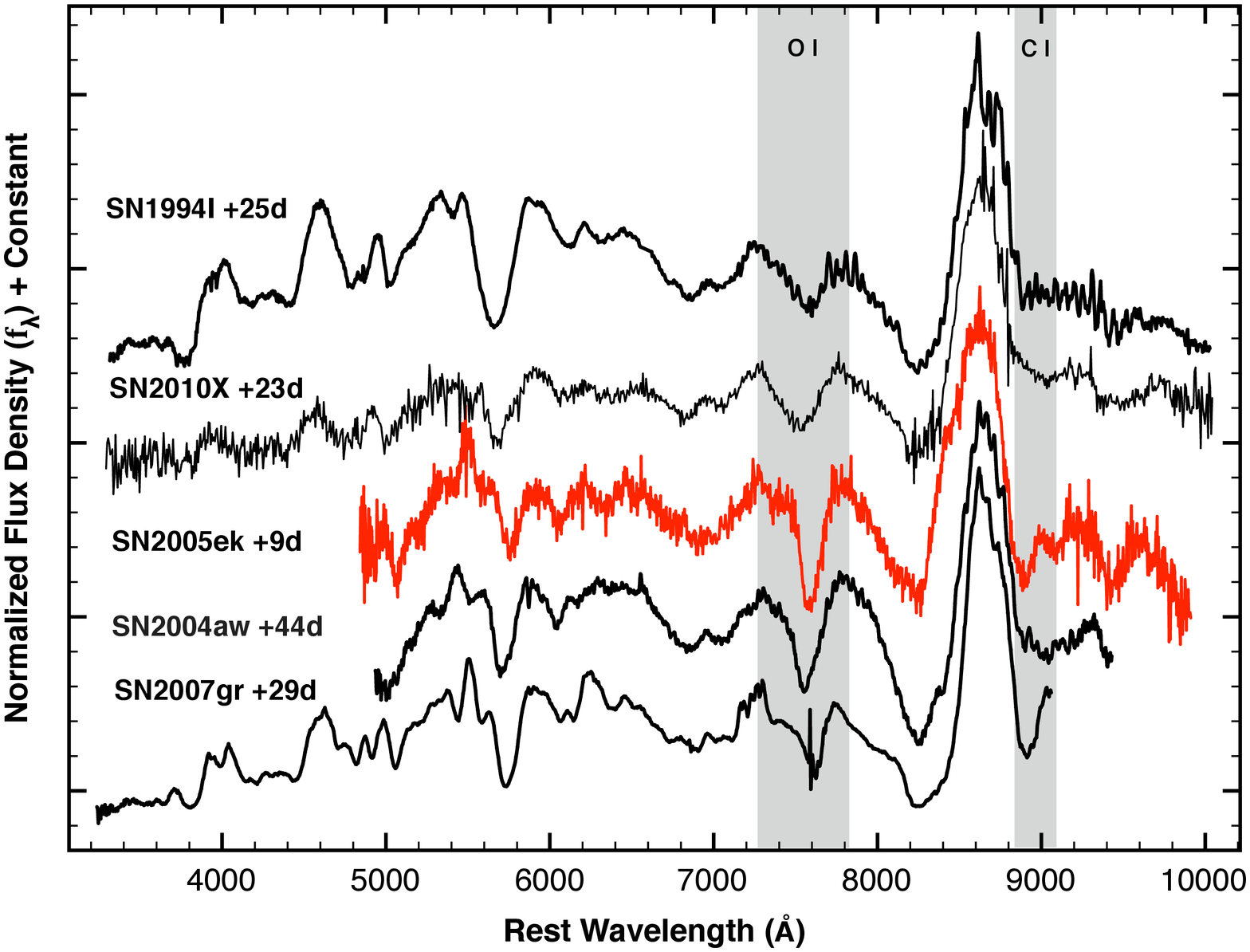}
\caption{\label{fig:Icspec} Comparison of SN\,2005ek (red) and other well-studied SN~Ic (black; SN\,1994I, SN\,2004aw, SN\,2007gr, and SN\,2010X).  Strong similarities are seen. \emph{top:} Near maximum light. The regions around \ion{C}{2} $\lambda$6582 and \ion{O}{1} $\lambda$7774 are shaded. \emph{bottom:} Transitional spectra. The region around \ion{O}{1} $\lambda$7774 and \ion{C}{1} $\lambda$9095 are shaded.}
\end{figure}

The spectra of SN\,2005ek shown in Figure~\ref{fig:SpecEvol} show considerable evolution over a short time period, and most closely resemble those of normal SN~Ic.  In Figure~\ref{fig:Icspec} we compare the maximum-light and transitional\footnote{``Transitional'' refers to the transition from optically thick to thin, which, in this case, we characterize by an increased prominence of the emission component of the Ca~II near-IR triplet)} spectra of SN\,2005ek to a set of SN~Ic (SN\,2010X, SN\,1994I, SN\,2007gr, and SN\,2004aw).  SN\,2005ek reaches the transitional phase much faster than the other events but the spectroscopic similarities at both epochs are clear. Of the events displayed in Figure~\ref{fig:Icspec}, only SN\,2010X decays on a timescale similar to that of SN\,2005ek.  The others possess $\Delta m_{15,R}$ values ranging from $\sim 1.4$ (SN\,1994I) to $\sim 0.4$ (SN\,2004aw).  

In this section, we examine the spectroscopic properties and evolution of SN\,2005ek utilizing two modeling techniques.  Initial line identifications and estimates of photospheric velocities were made with the spectral synthesis code SYN++ \citep{Thomas2011}\footnote{This is an updated version of SYNOW; https://c3.lbl.gov/es/ .}.  This analysis offers information about the ions present in a particular ionization state and spectral range, allowing one to cover a large parameter space with minimal time and computational resources.  In addition, we model a subset of the spectra using a one-dimensional Monte Carlo radiative transport code developed for SN outflows \citep{Mazzali1993,Lucy1999,Mazzali2000}, which allows quantitative assessments of ion abundances to be made.  Using the results of both techniques, we compare several distinctive spectroscopic features of SN\,2005ek to those of other SN~I and comment on the consequences for explosive nucleosynthesis.

\subsection{SYN++ Evolution and Photospheric Velocities} 

SYN++ is a parameterized spectral synthesis code which allows empirical fitting of SN spectra without {\it a priori} assumptions of the ejecta's density and composition structure.  It operates under the assumption of spherical symmetry, homologous expansion (radius proportional to velocity), a sharp photosphere, and a pseudo-blackbody continuum level.  Line formation is due to pure resonant scattering (treated using the Sobolev approximation) and Boltzmann statistics are utilized to determine the relative line strengths for a given ion. For more details, see \citet{Branch2002}, and \citet{Thomas2011}.  

\begin{figure}[!ht]
\includegraphics[width=\columnwidth]{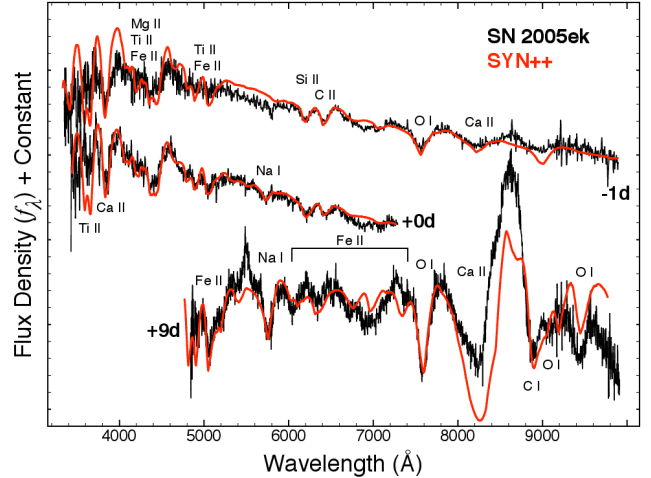}
\caption{\label{fig:Jerod2} SYN++ model fits to the $-1$ day, maximum-light, and $+9$ day spectra of SN\,2005ek (red lines). Observed spectra are shown in black.  Major spectroscopic features are labeled.}
\end{figure}

In Figure~\ref{fig:Jerod2} we show three representative SYN++ fits, covering the evolution of SN\,2005ek between $-1$ and +9 days. Major spectroscopic features are labeled. In all cases, the excitation temperature was set to 10,000 K and we chose an exponential form for the optical depth profile.  The photospheric velocity used in these fits ranges from $\sim 9000$ km s$^{-1}$ (day $-1$) to $\sim 7000$ km s$^{-1}$ (day +9).

\begin{figure}[!ht]
\includegraphics[width=\columnwidth]{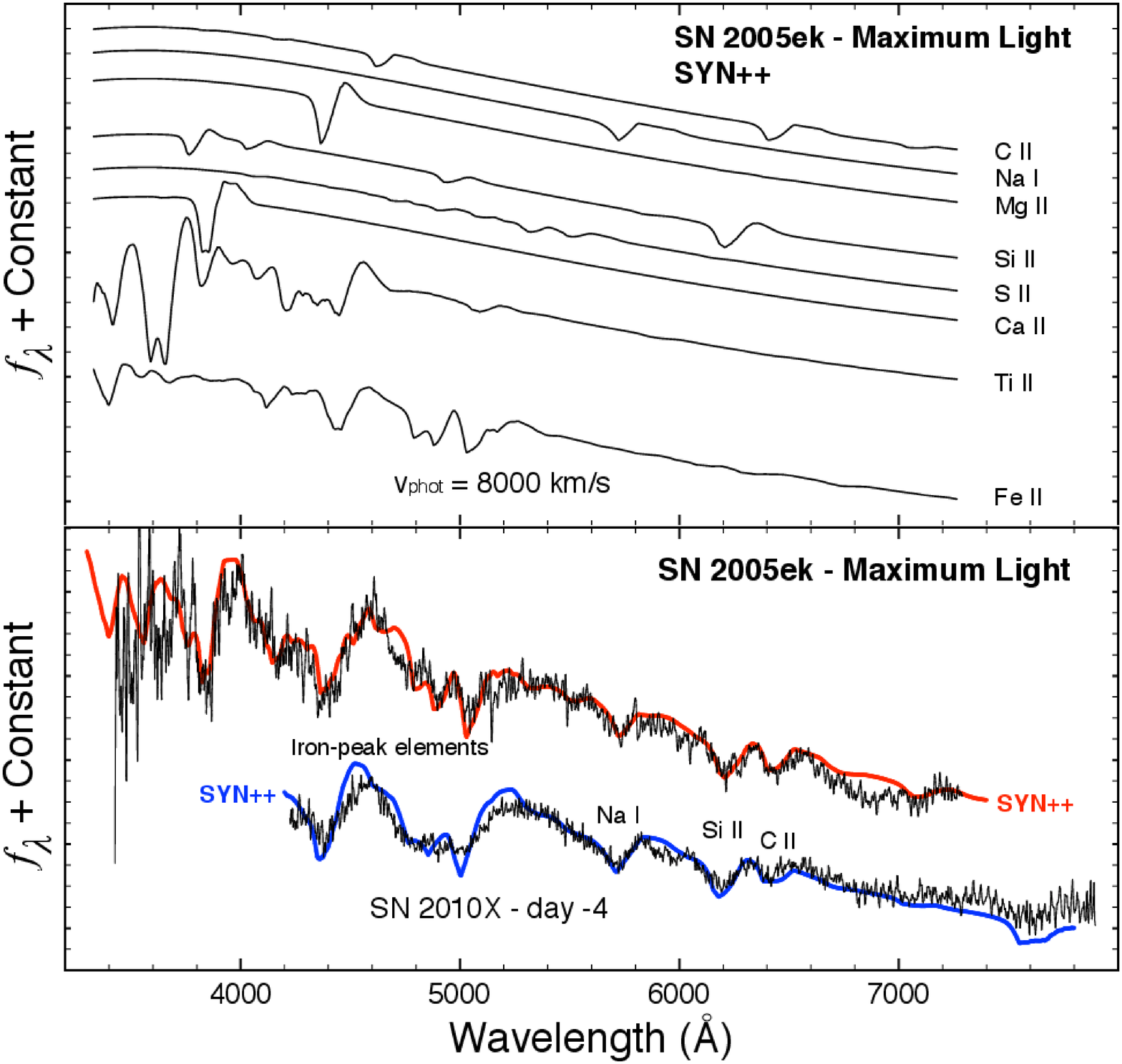}
\caption{\label{fig:Jerod1} \emph{top:} SYN++ model for the maximum-light spectrum of SN\,2005ek, separated by ion.  \emph{bottom:} Full model (red) along with a similar model constructed for the $-4$ day spectrum of SN\,2010X (blue). Observed spectra are shown in black.}
\end{figure}

The near-maximum-light spectra of SN\,2005ek can be modeled with a combination of \ion{O}{1}, \ion{C}{2}, \ion{Mg}{2}, \ion{Si}{2}, \ion{Ca}{2}, \ion{Ti}{2}, and some \ion{Fe}{2} at 8000--9000 km s$^{-1}$.  S~II is also included, although evidence for it is very weak. Between the $-1$ day and maximum-light models \ion{Na}{1} was added to describe the feature near 5700 \AA.  In Figure~\ref{fig:Jerod1} we present the individual ion components of the maximum-light model.  Also displayed (blue, lower panel) is a model constructed from the same set of ions for the $-4$ day spectrum of SN\,2010X.  This highlights the similarities between the spectra of SN\,2005ek and SN\,2010X and demonstrates that our fitting scheme is equally applicable to both events.   

Despite growth in the emission component of the \ion{Ca}{2} near-IR triplet, the +9 day spectrum of SN\,2005ek still shows a partial environment of resonant-line scattering, indicating it can be approximately modeled with SYN++.  By this epoch, the \ion{Si}{2} and \ion{C}{2} features found near maximum light have already faded. A majority of the features can be attributed to \ion{Fe}{2} and \ion{O}{1}, along with \ion{Na}{1}, \ion{Ca}{2}, and a decent fit to \ion{C}{1} $\lambda$9095 (consistent with the presence of \ion{C}{2} at earlier epochs) at $\sim 7000$ km s$^{-1}$.   SYN++ fits to the intermediate epochs are consistent with the decrease in velocity, increase in prominence of \ion{Fe}{2} and \ion{Ti}{2} features, and decrease in prominence of the \ion{C}{2} and \ion{Si}{2} features between the maximum-light and +9 day spectra.

\subsection{Abundance Modeling}\label{sec:Paolo}

To obtain quantitative estimates of the elemental abundances present in the ejecta of SN\,2005ek, we also model the $-1$ day, maximum-light, and +9 day spectra with the one-dimensional Monte Carlo radiation transport code described by \citet{Mazzali1993}, \citet{Lucy1999}, and \citet{Mazzali2000}.  The code assumes spherical symmetry and that radiation is emitted as a blackbody at a lower boundary (a pseudo-photosphere). The SN ejecta are defined by a run of density vs.\ velocity (an ``explosion model'' ) and a depth-dependent set of abundances.  Energy packets are allowed to interact with the ejecta gas via excitation processes and electron scattering. The state of the gas is computed according to that of the radiation using a lambda iteration and adopting the modified nebular approximation \citep{Mazzali1993}, while the emerging spectrum is computed by formally solving the transfer equation in a final step \citep{Lucy1999,Mazzali2000}. This method has been successfully applied to both SN~Ia (e.g., \citealt{Mazzali2008}) and SN~Ib/Ic (e.g., \citealt{Sauer2006}). 

In order to model the rapidly evolving spectra of SN\,2005ek we first had to establish a reasonable explosion model. The fast light curve and moderate velocities indicate a small ejecta mass. We experimented with different possibilities, and found that a model with mass $\sim 0.3$ M$_\odot$ and $E_K \approx 2.5 \times 10^{50}$ ergs provides a reasonable match to the evolution of the spectra.  The distribution of density with velocity resembles a scaled-down W7 model \citep{Nomoto1984}.  

Using this model, and assuming a rise time of 14 days, we were able to reproduce the spectroscopic evolution of SN\,2005ek (Fig.~\ref{fig:Paolo}).  As the photosphere recedes inward, each model epoch constrains a larger amount of the ejected material. With spectral coverage out to +9 days we are able to probe the outer $\sim 0.2$ M$_{\odot}$ ($\sim 66$\%) of the ejecta.  

The results are qualitatively consistent with the SYN++ line identifications given above. However, abundances do not necessarily correlate with line strength.  Notably, the strength of the \ion{O}{1} $\lambda$7774 line requires a high oxygen abundance, so that the composition is dominated by oxygen ($\sim 80$\%).  Fe is present and is responsible for the absorption near 5000\,\AA, while Ti and Cr are also important for shaping the spectrum near 4000\,\AA, but small abundances are sufficient for this as well as for Ca, despite the strength of the near-IR triplet absorption.  Smaller fractions of Mg, C, and Si are also present; Mg~II lines leave a strong imprint in the earlier spectra, especially in the red (features at 8900\,\AA\ and 7600\,\AA, the latter in a blend with the stronger O~I line), but also in the blue (4481\,\AA, which is the stronger contributor to the absorption near 4000\,\AA). 

The composition shows little or no variation between the three epochs. The ejecta are dominated by oxygen ($\sim 86$\%), with intermediate mass elements such as C, Mg, Si, and limited amounts of S and Ca contributing $\sim 13.5$\% and iron-peak elements (Fe, Ni, Cr, Co, Ti) contributing only $\sim 0.5$\%.

\subsection{Notable Spectroscopic Features}
The ions utilized in the models above are typical of SN~I.  However, several features warrant further discussion in the context of SN\,2005ek.

\subsubsection{Carbon Features}
The \ion{C}{2} $\lambda$6582 feature near 6400 \AA\ is noticeably prominent in the near-maximum-light spectra of SN\,2005ek. Its strength is comparable to the \ion{Si}{2} $\lambda$6355 feature.  This is due in part to the relative weakness of the \ion{Si}{2} feature, which, in the one-dimensional Monte Carlo models presented above, is caused by a combination of the low overall Si abundance ($\sim 2$\%) and the relatively low-density environment. However, the strength of the \ion{C}{2} feature is unusual in its own right, and it is still present several days post-maximum (Fig.~\ref{fig:SpecEvol}).

\begin{figure}[!ht]
\includegraphics[width=\columnwidth]{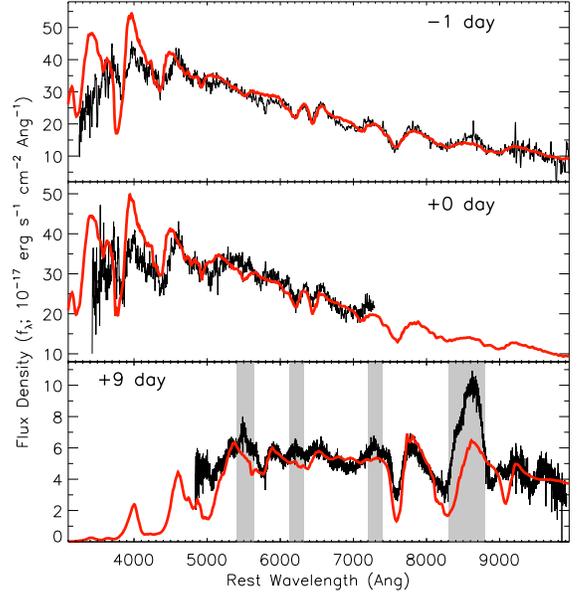}
\caption{\label{fig:Paolo} One-dimensional Monte Carlo radiation transport models for the $-1$ day (top panel), maximum-light (middle panel), and $+9$ day (lower panel) spectra of SN\,2005ek. Observed spectra are shown in black, models in red.  Regions showing excess with respect to the $+9$ day model, which may represent the onset of nebular features, are shaded (grey).}
\end{figure}

In SN~Ia, carbon features trace the distribution of unburned material from the carbon-oxygen (C-O) white dwarf.  \ion{C}{2} features are observed in $\sim$30\% of SN~Ia although they are rarely either this strong or after maximum light \citep{Howell2006,Thomas2007,Parrent2011,Silverman2012b,Folatelli2012}.  In general, the detection of carbon is also considered to be unusual in SN~Ic. However, all the spectra shown in Figure~\ref{fig:Icspec} also show evidence for a notch in this region (highlighted in grey), exemplifying the spectroscopic similarity of these objects.  This feature was specifically identified as C~II in SN\,2007gr \citep{Valenti2008} and SN\,2004aw \citep{Taubenberger2006}, while it is more debated in SN\,1994I (\citealt{Wheeler1994}) where features are broader and, hence, more blended.  In the bottom panel of Figure~\ref{fig:Icspec}, we also highlight the region around the feature we identify as \ion{C}{1} $\lambda$9095. A similar notch is seen in SN\,2007gr. Note that despite the strength of the \ion{C}{2} feature, the inferred carbon to oxygen abundance ratio is still small ($\sim$0.02).

\subsubsection{Iron-Peak Features}
In the $-1$ day spectrum of SN\,2005ek the broad feature between 4600 \AA\ and 5200 \AA, attributed mainly to \ion{Fe}{2}, is noticeably weaker than in the pre-maximum-light spectra of other SN~Ic (top panel of Fig.~\ref{fig:Icspec}).  The depth of the feature does increase in later epochs (Fig.~\ref{fig:SpecEvol} and Fig.~\ref{fig:fastspec}). Attempting to replicate this feature with other iron-peak elements produces less satisfactory model spectra.   

The presence of a strong \ion{Ti}{2} feature around 4400 \AA\ is one of the distinctive spectroscopic features of the rapidly-declining SN\,1991bg-like SN~Ia.  Although \ion{Ti}{2} is also identified in the models above, the lack of a strong \ion{Si}{2} absorption line (another distinctive feature of SN\,1991bg-like SN) distinguishes SN\,2005ek from these events. 

\subsubsection{Helium Features}
None of the models presented above include helium.  However, although its presence is not required, our data do not completely rule out its presence.  Due to the similarity in wavelength of \ion{He}{1} ($\lambda$5875, $\lambda$6678) to \ion{Na}{1} $\lambda$5890 and \ion{C}{2} $\lambda$6582, some ambiguity between ions can result (e.g., \citealt{Kasliwal2010}). In our case, the lack of features between 5200 \AA\ and 5800 \AA\ in the $-1$ day spectrum (see Fig.~\ref{fig:Icspec}) would require any such helium features to have increased in strength after maximum light. This would not be inconsistent with our abundance models, where we required an increase in sodium abundance to model the \ion{Na}{1} feature in the +9 day spectrum. However, \ion{He}{1} is difficult to excite \citep{Mazzali1998,Dessart2012}, especially in relatively cool radiation fields like those of SN\,2005ek, and we have no direct evidence for helium emission.  We find that the lack of helium in our fits does not require the presence of aluminum as implied by \citet{Kasliwal2010}.  Rather, a majority of the SN features can be reproduced with a combination of \ion{Fe}{2} and \ion{Na}{1}.

\subsubsection{Nebular Features}
The +9 day spectrum of SN\,2005ek appears to be transitional between the photospheric and nebular phases, with the Ca~II near-IR feature significantly influenced by net emission.  In addition, there are slight excesses with respect to both the SYN++ and Monte Carlo models near 5500 \AA, 6300 \AA, and 7300 \AA\ (shaded regions, bottom panel, Fig.~\ref{fig:Paolo}) which may be due to the emergence of [O~I] $\lambda$5577, [O~I] $\lambda \lambda$6300, 6364, and [Ca~II] $\lambda\lambda$7291, 7324 in emission.  This may represent the earliest onset of nebular features in a SN~I observed to date.

\subsection{Comparison to SN\,2002bj}

\begin{figure}
\includegraphics[width=\columnwidth]{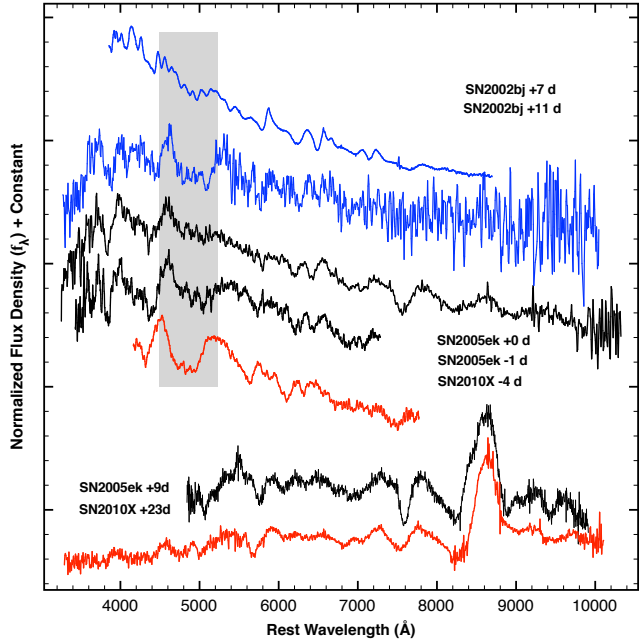}
\caption{\label{fig:fastspec} Comparison of SN\,2005ek, SN\,2010X, and SN\,2002bj spectra.  \emph{top:} ``Early''-time spectra. \emph{bottom:} ``Late''-time spectra. No spectra from a similar phase are available for SN\,2002bj.}
\end{figure}

The decline rate of SN\,2005ek is most closely matched by SN\,2002bj and SN\,2010X and in Figure~\ref{fig:fastspec} we compare the spectra of these three objects. While SN\,2010X possesses very similar spectroscopic features at both early and late times (with moderately higher photospheric velocities), the spectroscopic similarities between SN\,2005ek and SN\,2002bj are less clear.  

\citet{Poznanski2010} model SN\,2002bj with SYNOW, finding that a majority of features can be fit with intermediate-mass elements (C, Si, S) and helium at the relatively low velocity of $\sim 4000$ km s$^{-1}$.  Notably absent from their fit were any iron-peak elements.  This is in stark contrast to the maximum-light spectra of SN\,2005ek where both iron and titanium play a significant role, and sulfur is not required.  Indeed, the first spectrum of SN\,2002bj (top, Fig.~\ref{fig:fastspec}) shows a significantly bluer continuum than any of the spectra obtained for either SN\,2005ek or SN\,2010X.  This spectrum was obtained seven days post-maximum.  By +11 days (when a second, lower quality, spectrum was obtained), SN\,2002bj appears to show a distinctly different morphology in the blue (Fig.~\ref{fig:fastspec}).  The continuum is more depressed in the region between 3000 \AA\ and 5500 \AA, with features that appear similar to those attributed to iron-peak elements in SN\,2005ek. This spectroscopic evolution is clearly distinct from that of SN\,2005ek and SN\,2010X.

\subsection{Consequences for Nucleosynthesis}\label{sec:nuc}

The ejecta of SN\,2005ek are dominated by oxygen.  Unlike the ``calcium-rich'' objects, such as SN\,2005E whose ejecta was $\sim 50$\% calcium \citep{Perets2010}, this does not necessarily imply an  unusual nucleosynthetic channel.  Oxygen-dominated ejecta are common in models of SN~Ic which are due to core collapse in a stripped C-O star.  In that situation, the large oxygen abundance is due to a combination of the initial composition along with partial carbon burning.  In contrast, producing oxygen-dominated ejecta via primarily helium burning (the mechanism invoked in the ``.Ia'' scenario) is not straightforward. \citet{Perets2010} use a one-zone model to examine the explosive nucleosynthetic outputs of He, C, and O mixtures at several temperatures. None of their helium-dominated trials yield oxygen-dominated ejecta.  

\section{Power Source and Explosion Parameters}\label{sec:explosion}

The optical light curves of normal SN~Ia and SN~Ib/Ic are powered mainly by the radioactive decay of $^{56}$Ni.  For SN\,2005ek, the change in light-curve slope between +20 and +40 days (Figure~\ref{fig:PseudoB}) points to radioactive decay as a possible power source. However, the initial decay timescale as well as the almost linear post-maximum decline are unusual. In this section we examine whether a $^{56}$Ni power source is consistent with the observations of SN\,2005ek described above.

For the purposes of simplified analytic models, the bolometric light curves of SN~Ib/Ic are usually divided into two regimes as follows. (a) The \emph{photospheric phase}, when the optical depth is high and the shape of the light curve is dependent both on the rate of energy deposition and the photon diffusion time scale \citep{Arnett1982}.  In normal SN~Ib/Ic, the effects of radiative transfer result in an initial post-maximum decline rate which is \emph{slower} than $^{56}$Ni $\rightarrow$ $^{56}$Co decay \citep{Drout2011}. (b) The \emph{nebular phase}, when the optical depth has decreased and the SN luminosity is determined by the instantaneous rate of energy deposition.  Normal SN~Ib/Ic enter this stage at a late epoch ($\gtrsim 60$ days;\citealt{Valenti2007}) when the dominant energy source is expected to be $^{56}$Co $\rightarrow$ $^{56}$Fe decay.  The late-time slope of SN~Ib/Ic light curves are well matched by this decay rate when the effects of incomplete gamma-ray trapping are included \citep{Clocchiatti1997,Valenti2007}.

\begin{figure}[!ht]
\includegraphics[width=\columnwidth]{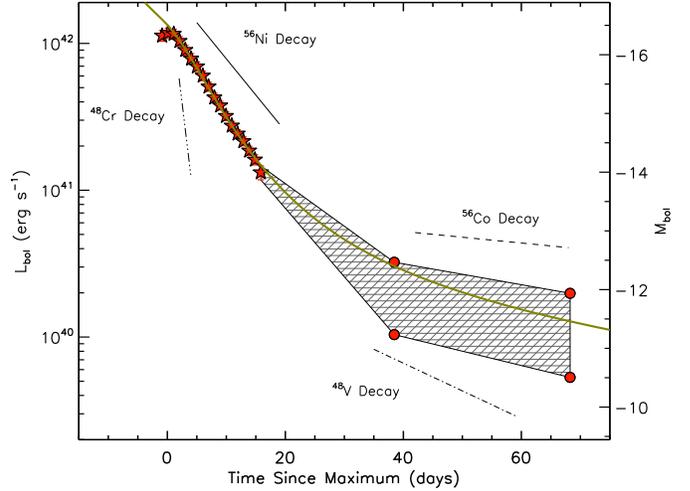}
\caption{\label{fig:Nickel} Radioactive models for the pseudo-bolometric light curve of SN\,2005ek.  Black lines show the decay rates for $^{56}$Ni, $^{56}$Co, $^{48}$Cr, and $^{48}$V, assuming full trapping of gamma-rays.  The gold curve shows the best-fit model described in the text, assuming $\sim 0.03$ M$_\odot$ of $^{56}$Ni, a luminosity which tracks the instantaneous energy input, and incomplete gamma-ray trapping.}
\end{figure}

The time when a SN will transition to the second of these two phases is determined in large part by the total ejecta mass and kinetic energy of the explosion.  In SN\,2005ek, both the small inferred ejecta mass ($\sim 0.3$ M$_\odot$) and the early onset of nebular spectroscopic features indicate that the assumption of optically thick ejecta may break down within a few days of maximum light, making the models of \citet{Arnett1982} inapplicable.  Further, we note that the early portion of the pseudo-bolometric light curve appears linear, and decays at a rate of 0.15 mag day$^{-1}$, comparable to the 0.12 mag day$^{-1}$ given by $^{56}$Ni $\rightarrow$ $^{56}$Co decay.  In Figure~\ref{fig:Nickel} we again plot the pseudo-bolometric light curve of SN\,2005ek.  Also shown are lines which describe the decay rate of $^{56}$Ni $\rightarrow$ $^{56}$Co and $^{56}$Co $\rightarrow$ $^{56}$Fe.

With this early decay rate as motivation, we construct a model for the entire post-maximum pseudo-bolometric light curve of SN\,2005ek based on the instantaneous rate of energy deposition from the $^{56}$Ni $\rightarrow$ $^{56}$Co $\rightarrow$ $^{56}$Fe decay chain. The model is similar to the nebular phase model of \citet{Valenti2007}, although we allow for incomplete trapping of the gamma-rays produced from $^{56}$Ni $\rightarrow$ $^{56}$Co decay.  One effect of incomplete trapping at this early phase is that the light curve should decay by a larger number of magnitudes before settling onto the $^{56}$Co tail.  This prediction is in good agreement with our comparison of SN\,2005ek to other SN~I in \S \ref{lcs} (see Fig.~\ref{fig:AppPan}).  

Under these assumptions, the luminosity of the SN can be modeled as (\citealt{Valenti2007}, \citealt{Sutherland1984}, \citealt{Cappellaro1997}; we use the notation of \citealt{Valenti2007})
\vspace{0.2 in}

\noindent$L (t) = S^{\rm Ni}(\gamma) +S^{\rm Co}(\gamma)+S^{\rm Co}_{e^+}(\gamma)+S^{\rm Co}_{e^+}({\rm KE})$,

\vspace{0.2 in}

\noindent where the four terms describe the energy due to gamma-rays from nickel decay, gamma-rays from cobalt decay, gamma-rays from the annihilation of positrons created in cobalt decay, and the kinetic energy of positrons created in cobalt decay, respectively.  These are given by

\vspace{0.2 in}

\noindent $S^{\rm Ni}(\gamma) = \mathrm{M}_{\mathrm{Ni}} \epsilon_{\mathrm{Ni}} e^{-t/\tau_{\mathrm{Ni}}} (1 - e^{-F/t^2})$ \\
$S^{\rm Co}(\gamma) = 0.81 \times \mathcal{E}_{\mathrm{Co}} (1 - e^{-(F/t)^2})$ \\
$S_{e^+}^{\rm Co}(\gamma) = 0.164 \times \mathcal{E}_{\mathrm{Co}} (1 - e^{-(F/t)^2}) (1 - e^{-(G/t)^2})$ \\
$S_{e^+}^{\rm Co}({\rm KE}) = 0.036 \times \mathcal{E}_{\mathrm{Co}} (1 - e^{-(G/t)^2})$,

\vspace{0.2 in}
\noindent
where

\vspace{0.2 in}

\noindent $\mathcal{E}_{\mathrm{Co}} = \mathrm{M}_{\mathrm{Ni}} \epsilon_{\mathrm{Co}}( e^{-t/\tau_{\mathrm{Co}}}-e^{-t/\tau_{\mathrm{Ni}}})$.

\vspace{0.2 in}

The incomplete trapping of gamma-rays and positrons is incorporated with the terms $(1 - e^{-(F/t)^2})$ and $(1 - e^{-(G/t)^2})$.  $F$ and $G$ are constants such that the gamma-ray and positron optical depths decrease by a factor proportional to $t^{-2}$ as expected for an explosion in homologous expansion \citep{Clocchiatti1997} and are functions of the total ejecta mass, kinetic energy, and density distribution of the ejecta \citep{Clocchiatti1997}.  Using this model with $M_{\rm Ni} = 0.03$ M$_\odot$, $F = 12.8$ days, and G $\approx 16.1F$ \citep{Valenti2007}, we find the gold curve shown in Figure~\ref{fig:Nickel} which matches the early decay rate of SN\,2005ek and is also consistent with our late-time constraints.  Adopting the parameterization of \citet{Valenti2007} where $F \approx 32~M_{\rm ej,\odot}/\sqrt{E_{K,51}}$, this value of $F$ implies a $M_{\rm ej,\odot}/\sqrt{E_{K,51}} \approx 0.4$.  Using the observed photospheric velocity near maximum ($\sim 8500$ km s$^{-1}$), we can break the degeneracy between $M_{\rm ej}$ and $E_K$ to yield explosion parameters of $M_{\rm ej} \approx 0.7$ M$_\odot$ and $E_K \approx 5.2 \times 10^{50}$ ergs.  These values are a factor of two larger than those used in our spectroscopic modeling ($M_{\rm ej} = 0.3$ M$_\odot$ and $E_K = 2.5 \times 10^{50}$ ergs; \S \ref{sec:Paolo}), but given the number of assumptions required to extract explosion parameters from this simplified analytic model the two are relatively consistent.  We adopt conservative estimates of the explosion parameters to be $M_{\rm ej} = 0.3$--0.7 M$_\odot$ and $E_K = 2.5$--5.2 $\times 10^{50}$ ergs.

It has been suggested \citep[e.g.][]{Shen2010} that other radioactive decay chains such as $^{48}$Cr $\rightarrow$ $^{48}$V $\rightarrow$ $^{48}$Ti, which possess shorter decay times than $^{56}$Ni $\rightarrow$ $^{56}$Co $\rightarrow$ $^{56}$Fe, may contribute to the luminosity of rapidly evolving events.  In Figure~\ref{fig:Nickel} we also include lines that represent the decay rates of $^{48}$Cr $\rightarrow$ $^{48}$V and $^{48}$V $\rightarrow$ $^{48}$Ti.  The rapid $^{48}$Cr decay time ($\tau_{\rm Cr} = 1.3$ days) implies that by a few days post-explosion the power input should already be dominated by $^{48}$V $\rightarrow$ $^{48}$Ti decay.  Although photon diffusion likely plays a role at very early times, it is difficult to reconcile this power source with the change in light-curve slope observed between +20 and +40 days.  In addition, in order to fit both of our late-time luminosity constraints with $^{48}$V decay, we would require nearly full gamma-ray trapping ($\sim 0.05$ mag day$^{-1}$, five times steeper than $^{56}$Co), which is inconsistent with our low derived ejecta mass. Thus, although we cannot completely rule out some (especially early) contributions from other radioactive decay chains, we find that our observations are consistent with SN\,2005ek being powered by the radioactive decay of $\sim 0.03$ M$_\odot$ of $^{56}$Ni.  We summarize this and our other inferred explosion parameters in Table~\ref{tab:ExpPara}. 

\section{Host Galaxy: UGC 2526}\label{sec:HostProps}

\subsection{Global Properties}

\begin{figure}
\includegraphics[width=\columnwidth]{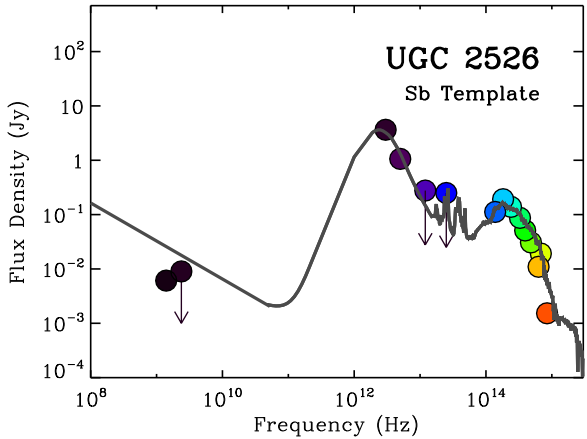}
\caption{\label{fig:hostsed} SED for UGC\,2526 (circles) and the best-fit Sb galaxy model (grey line).}
\end{figure}

SN\,2005ek exploded in the outskirts of UGC 2526, an edge-on spiral galaxy of morphology Sb.  In Figure~\ref{fig:hostsed} we show the SED of UGC 2526, which was compiled from the SDSS \citep{Ahn2012} and IRAS \citep{MD2005} catalogs and supplemented with upper limits from our radio observations described in \S \ref{sec:Obs}.  Also shown is an Sb model template from the SWIRE database \citep{Silva1998}. Using this template, we derive a star-formation rate (SFR) of $\sim 2$--5 M$_\odot$ yr$^{-1}$ \citep{Yun2002,Kennicutt1998}.  SN\,2010X, SN\,2002bj and SN\,1885A also exploded in star-forming galaxies.

UGC\,2526 has a low radio luminosity when compared to the Sb template which provides a best fit at other wavelengths. According to \citet{Chakraborti2012} this can be explained if electrons responsible for producing the radio synchrotron emission undergo significant inverse-Compton losses. Synchrotron and inverse-Compton losses are proportional to the energy density in magnetic fields and seed photons, respectively. The energy density of a characteristic Milky-Way-like magnetic field of $\sim 5~\mu$G is $\sim 10^{-12}$ ergs cm$^{-3}$. We estimate that the bolometric luminosity of the host galaxy contributes an energy density of $\sim 3 \times 10^{-12}$ ergs cm$^{-3}$ which could lead to significant inverse-Compton losses.

We measure the metallicity of UGC\,2526 using our host-galaxy spectrum centered on the galaxy nucleus (\S \ref{sec:Obs}).  H$\alpha$ and [N~II] line fluxes were measured using the Markov Chain Monte Carlo (MCMC) technique described by \citet{Sanders2012a} and, using the relations of \citet{Pettini2004}, we find the metallicity of the UGC 2526 to be 12 $+\log{\rm (O/H)}_{\rm PP04N2} = 8.79 \pm 0.06$.  This value is approximately solar\footnote{12 $+\log{\rm (O/H)}_{\rm solar} = 8.69$ on the PP04N2 scale}.  This metallicity measurement may be affected by absorption from the underlying stellar population, although H$\alpha$ should be relatively less affected than H$\beta$, resulting in a slight overestimation of the actual metallicity. For typical SN host galaxies the correction factor is on the order of 0.05 dex and in extreme cases $\sim 0.2$ dex.  Additionally, the relatively high [N~II] contribution suggests that the galaxy may be weakly active \citep[probably a LINER; e.g.,][]{HFS97}.

\subsection{Explosion-Site Properties}

The explosion site of SN\,2005ek is offset nearly 30 kpc ($1.5'$) in projection from the center of UGC 2526, which possesses a major diameter of $D_{25} \approx 69$ kpc. This places SN\,2005ek at the extreme high end of the distribution of offsets seen for all SN subtypes \citep{Prieto2008,Kelly2011}, which is especially notable because SN~Ib/Ic typically exhibit smaller offsets than SN~II and SN~Ia.  This large offset, coupled with the observation of metallicity gradients in many spiral galaxies \citep{Zaritsky1994}, implies that the explosion-site metallicity is likely lower than the value we measured in the galaxy nucleus. In contrast, SN\,2010X, SN\,2002bj, and SN\,1885A all exploded with low projected offsets \citep{Kasliwal2010,Poznanski2010,Perets2011}. 

The lack of nebular emission lines in the explosion-site spectrum allows us to place a strict limit on the amount of star formation within the $1''$ slit ($\sim 0.3$ kpc at the distance of UGC 2526).  We measure a $3\sigma$ upper limit on the H$\alpha$ line flux of $3.3\times 10^{37}$ ergs s$^{-1}$, which corresponds to an upper limit on the local SFR of $2.6 \times 10^{-4}$ M$_\odot$ yr$^{-1}$ (Eq.\ 2, \citealt{Kennicutt1998}).  While this value is an order of magnitude below the mean H$\alpha$ flux measured for H~II regions associated with core-collapse supernovae in the sample of \citet{Crowther2012}, negligible H$\alpha$ flux at the explosion site of a core-collapse supernova is not unprecedented. Only 21 of 39 supernovae in the \citet{Crowther2012} sample show evidence for an associated H~II region. 

However, a much higher fraction of SN~Ib/Ic are associated with H~II regions compared to SN~II ($70 \pm 26$\% versus $38 \pm 11$\%; \citealt{Crowther2012}). This trend may reflect the fact that massive stars with $M \lesssim 12$ M$_\odot$ (which are expected to explode as SN~II) have lifetimes longer than the roughly 20 Myr lifetimes of giant \ion{H}{2} regions.  Thus, the lack of observed \ion{H}{2} emission makes the explosion site of SN\,2005ek more comparable to those of SN~II and is consistent with a progenitor older than $\sim 20$ Myr.

\section{Rates} \label{sec:rates}
  
SN\,2005ek was discovered as part of the LOSS survey, which achieved a high level of completeness.  From the LOSS data, \citet{Li2011} constructed volume-limited samples of Type Ia and core-collapse supernovae out to distances of 80 Mpc and 60 Mpc, respectively, which were used to derive relative rates for various SN subtypes.  Although SN\,2005ek was not included in this original analysis\footnote{Its distance of 66 Mpc and classification as a SN~Ic placed it just slightly outside the relevant sample volume (60 Mpc for core-collapse SN).}, we can use the LOSS data to obtain a rough estimate for the relative rate of SN\,2005ek-like transients. SN\,2002bj was also discovered by the LOSS survey, and we derive two sets of rates below, one of which assumes SN\,2002bj and SN\,2005ek are members of the same class of objects.

The LOSS volume-limited SN~Ia sample contains 74 objects and is 99\% complete to a distance of 80 Mpc.  In order to estimate the incompleteness correction for SN\,2005ek, we examine the correction factors for SN\,2002dk and SN\,2002jm, two SN\,1991bg-like objects with peak magnitudes similar to that of SN\,2005ek.  \citet{Li2011} find that both objects are $\sim 97$\% complete to a distance of 80 Mpc.  This correction factor is based on a combination of peak magnitude and light-curve shape, and should therefore be taken as an upper limit for the completeness factor of the rapidly declining SN\,2005ek.  If the completeness factor of SN\,2005ek lies between 50\% and 100\%, we may estimate that the rate of such transients is $\sim 1$--2\% of the SN~Ia rate.  If we include SN\,2002bj in the same category of objects as SN\,2005ek, this rate rises to $\sim 2$--3\% of the SN~Ia rate.  These rates should be taken as lower limits, and are consistent with those estimated by \citet{Poznanski2010} and \citet{Perets2011}.  It should also be noted that Poisson errors in this small number regime are large.

\section{Possible Progenitor Channels} \label{sec:theories}

The observations and analysis presented above allow us to examine several possible progenitor models for SN\,2005ek.  SN\,2005ek shows a very rapid post-maximum decline, a peak luminosity of $\sim 10^{42}$ ergs s$^{-1}$, colors which redden with time, and photospheric velocities which evolve from $\sim 9000$ km s$^{-1}$ near maximum to $\sim 7000$ km s$^{-1}$ at +9 days.  Spectroscopic modeling reveals a small ejecta mass (0.3--0.7 M$_\odot$) which is predominantly oxygen ($\sim 85$\%), with smaller amounts of other intermediate-mass elements (Mg, C, Si) and an explosion kinetic energy of (2.5--5.0) $\times 10^{50}$ ergs.  The pseudo-bolometric light curve is consistent with an explosion powered by $\sim 0.03$ M$_\odot$ of $^{56}$Ni, assuming a non-negligible fraction of the gamma-rays escape at early times.  This assumption is consistent with our low inferred ejecta mass and the emergence of nebular features at only 9 days post maximum light.  Finally, both the large offset and low level of H$\alpha$ emission from the explosion site of SN\,2005ek are consistent with a progenitor older than $\sim 20$ Myr.  

A robust progenitor model should be able to reproduce all of the above properties.  In addition, if one accepts that the spectroscopic and compositional similarities between SN\,2005ek and other normal SN~Ic imply that they should have a common class of progenitors, the model should be capable of producing a variety of observed decline rates and ejecta masses.  Possible progenitor channels can be divided into two classes: those involving a white dwarf (WD) or neutron star (NS) and those involving a massive star.  We examine both below.

\subsection{Degenerate Objects}\label{sec:degen}

Explosion models involving degenerate objects make attractive models for faint, rapidly evolving transients, as they naturally predict small ejecta masses.  In addition, the old stellar environment of SN\,2005ek is consistent with a progenitor system containing at least one degenerate object.  Here we discuss several specific scenarios in the context of SN\,2005ek: the accretion-induced collapse of a WD, a WD-NS or NS-NS merger, and the detonation of a helium shell on a low-mass WD.

\subsubsection{Accretion-Induced Collapse (AIC)}

Under certain circumstances, when an accreting WD nears the Chandrasekhar mass, electron capture may occur in its core, causing it to collapse to a NS rather than undergo a thermonuclear explosion \citep{Nomoto1991}.  Modern simulations suggest that the subsequent bounce and neutrino-driven wind can lead to the ejection of a small amount of material, producing a weak, rapidly evolving transient powered by radioactive decay (e.g., \citealt{Metzger2009,Darbha2010,Fryer1999,Fryer2009}). The observational properties of these AIC transients should vary if the AIC is caused by the merger of two WDs rather than the collapse of a single degenerate object.

The single-degenerate case is unlikely to produce a transient similar to SN\,2005ek.  The simulations of \citet{Dessart2006} predict ejecta masses and explosion energies of $\sim 10^{-2}$ M$_\odot$ and $\sim 10^{49}$ ergs, respectively, an order of magnitude below those inferred for SN\,2005ek.  This is evident in the far-left panel of Figure~\ref{fig:theories}, where the theoretical light curves of Darbha et al.\ (2010; black lines) are significantly faster and fainter than those of SN\,2005ek (red stars).  In addition, the predicted velocities are on the order of $0.1c$, and a majority of the ejecta is likely processed to nuclear statistical equilibrium, implying a dearth of intermediate-mass elements \citep{Darbha2010,Metzger2009,Fryer1999}.

An AIC due to the merger of two WDs may be ``enshrouded'' \citep{Metzger2009} by $\sim 0.1$ M$_\odot$ of unburned material left in a remnant disk \citep{Yoon2007}.  This material will be shock heated by the ensuing explosion, potentially synthesizing intermediate-mass elements, as well as slowing the initially rapid ejecta velocity.  However, models of \citet{Fryer2009} suggest that this heating may lead to a transient which peaks in the UV bands, inconsistent with our observations of SN\,2005ek.

\subsubsection{WD-NS/NS-NS Merger}

\begin{figure*}[!ht]
\begin{centering}
\includegraphics[width=0.98\textwidth]{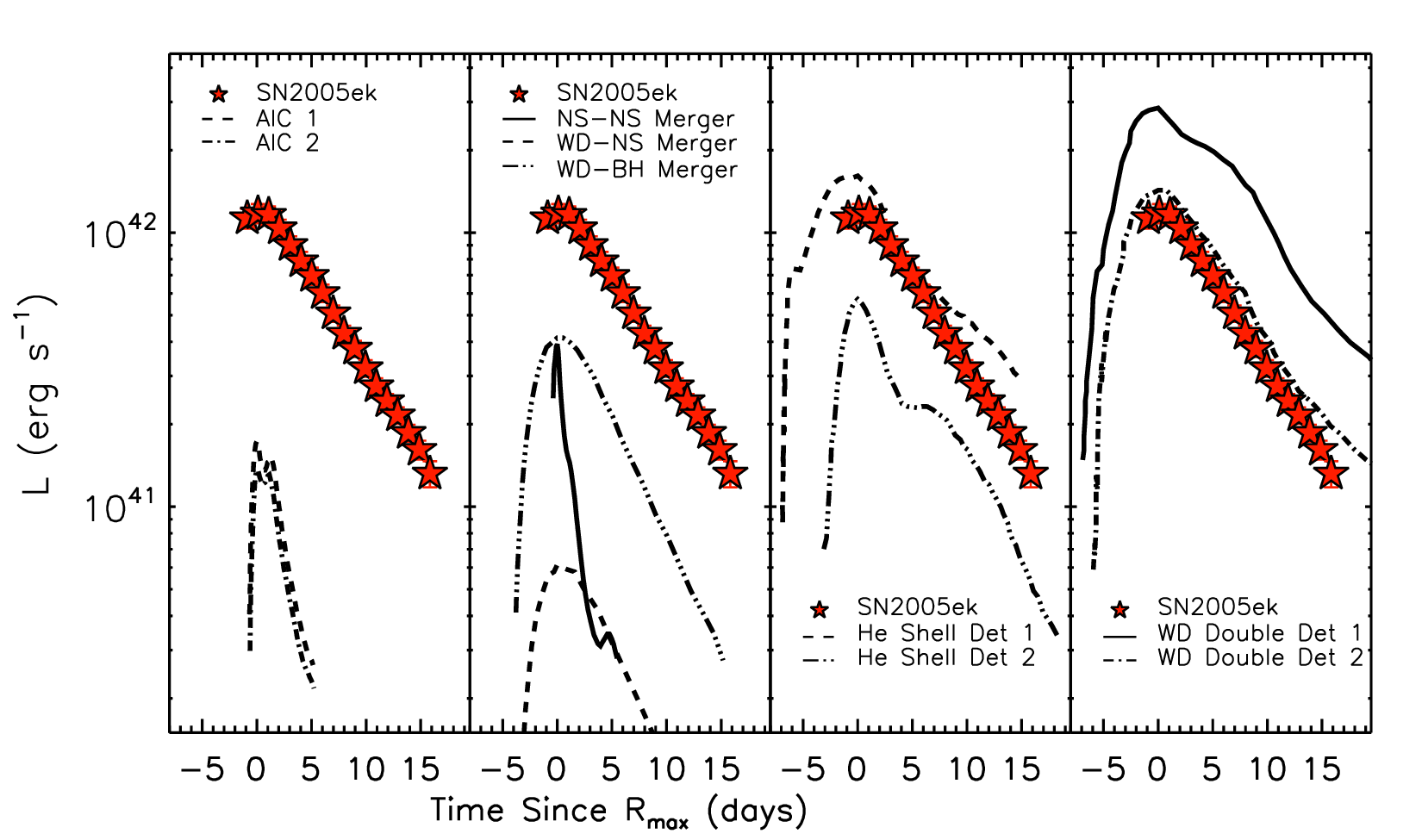}
\caption{\label{fig:theories} Comparison of the pseudo-bolometric light curve of SN\,2005ek (red stars) to theoretical light curves. 
\emph{far left:} Two AIC models from \citet{Darbha2010}. Both models are calculated for $M_{\rm ej} = 10^{-2}$ M$_\odot$. 
\emph{middle left:} The NS-NS merger from \citep{Metzger2010}, the WD-BH merger from \citep{Metzger2012}, and the WD-NS merger from \citep{Metzger2012}. 
\emph{middle right:}  Two ``.Ia'' models from \citep{Shen2010}. Models 1 and 2 represent the detonation of a 0.1 M$_\odot$ He shell on a 1.0 M$_\odot$ WD and a 0.05 M$_\odot$ He shell on a 1.2 M$_\odot$ WD, respectively.
\emph{far right:} Two edge-lit double-detonation models from \citet{Sim2012}.  Models 1 and 2 represent a 0.21 M$_\odot$ He shell on a 0.58 M$_\odot$ and 0.45 M$_\odot$ WD, respectively.
}
\end{centering}
\end{figure*}

Both NS-NS and NS-WD mergers have also been theorized to produce faint optical transients.  In the former case, r-process nucleosynthesis is thought to occur during the ejection of neutron-rich tidal tails, yielding a rapidly evolving transient with peak luminosities between $10^{41}$ and $10^{42}$ ergs s$^{-1}$ (\citealt{Metzger2010,Roberts2011}; solid line, center left panel of Figure~\ref{fig:theories}). However, similar to the single-degenerate AIC scenario described above, the ejecta masses are lower ($\sim 10^{-2}$ M$_\odot$), ejecta velocities are higher ($\sim 0.1c$) and the nucleosynthetic yields are inconsistent (mainly r-process elements) with our observations of SN\,2005ek.  In addition, using improved r-process opacities \citet{Barnes2013} find that these transients may be fainter, longer lived, and significantly redder than previously hypothesized.

In contrast, the tidal disruption of a WD by a NS or BH yields a set of explosion parameters at least broadly consistent with those observed for SN\,2005ek.  By examining the evolution of and nucleosynthesis within the accretion disk formed during the disruption \citet{Metzger2012} produce a set of models with ejecta masses between 0.3 and 1.0 M$_\odot$, ejecta velocities between 1000 and 5000 km s$^{-1}$, synthesized Nickel masses between $10^{-3}$ and $10^{-2}$ M$_\odot$, and peak luminosity between $10^{39}$ and $10^{41.5}$ ergs s$^{-1}$.  The disruption of a larger WD yields a transient with a larger ejecta mass, expansion velocity, and nickel mass. Two example light curves are shown as dashed lines in the middle-left panel of Figure~\ref{fig:theories}.  Although the curves fall slightly below our observations (a consequence of the slightly lower inferred nickel mass) the shape is reproduced.  In addition, because the outer layers of the disk do not burn to nuclear statistical equilibrium, the final ejecta composition is at least qualitatively consistent with our results for SN\,2005ek (mainly O, C, Si, Mg, Fe, and S), although the exact compositional fractions may not be reproduced.  In particular, in order to synthesize enough $^{56}$Ni to power SN\,2005ek, the current models would require a total ejecta mass greater than 1.0 M$_\odot$.  

\subsubsection{Helium-Shell Detonation}

Finally, we examine the detonation of a helium shell on the surface of a WD, a model for which numerous theoretical light curves and spectra have been produced \citep[e.g.,][]{Woosley1986,Shen2010,Fryer2009,Waldman2011,Sim2012} and which has been invoked to explain a number of unusual recent transients (e.g., SN\,2005E, \citealt{Perets2010}; SN\,2002bj, \citealt{Poznanski2010}; SN\,2010X, \citealt{Kasliwal2010}; SN\,1885A \citealt{Chevalier1988}). The term ``.Ia,'' which is often associated with this explosion mechanism, was used by \citet{Bildsten2007} to describe the specific case where the detonation occurs after mass transfer within an AM~CVn binary system.  In this case, the predicted helium-shell mass at the time of detonation is relatively small ($\lesssim 0.1$ M$_\odot$), which leads to a faint ($-15 > M_R > -18$ mag) and rapidly evolving transient.  In the center-right panel of Figure~\ref{fig:theories}, we show two ``.Ia'' model light curves from \citet{Shen2010}.  SN\,2005ek falls comfortably between the two.

However, despite the similarity in light-curve morphology, the ``.Ia'' model in its basic form fails to reproduce the inferred abundances of SN\,2005ek.  The detonation of a predominately helium shell should yield ejecta dominated by calcium, iron-peak elements (especially titanium), and unburned helium, with a notable absence of other intermediate-mass elements \citep{Shen2010, Perets2010}.  As discussed in \S \ref{sec:nuc}, it is not straightforward to produce a high oxygen abundance from helium burning, and although calcium and titanium features are present in the spectra of both SN\,2005ek and SN\,2010X, these do not correspond to high abundances.

One possible reconciliation of these issues is if the helium-shell detonation triggers a second detonation within the C-O WD.  Such ``double-detonation'' scenarios in solar-mass WDs have been thoroughly investigated as a possible explosion mechanism for normal SN~Ia.  However, \citet{Sim2012} extend this analysis to low-mass WDs, considering both core-compression and edge-lit secondary detonations.  In the former case significant amounts of iron-peak elements are synthesized and a much more slowly evolving light curve is produced.  However, in the latter case the main modification to the observable parameters is the production of additional intermediate-mass elements. Depending on the precise abundances synthesized and amount of unburned C-O material ejected, it is possible that this scenario can explain most of the observational properties of SN\,2005ek. In the far-right panel of Figure~\ref{fig:theories} we show two edge-lit models from \citet{Sim2012}, demonstrating that they are capable of reproducing the morphology of SN\,2005ek.

\subsection{Massive Stars}

In our above discussion, we found that both the edge-lit double detonation of a low mass WD and the tidal disruption of a WD by a NS could potentially explain the observed properties of SN\,2005ek.  However, as discussed in \S \ref{sec:SpecModel}, SN\,2005ek shows remarkable spectroscopic similarity to a number of normal SN~Ic.  If one assumes a single set of progenitors for these objects, and accepts their inferred ejecta masses (between 1 and 10 M$_\odot$; \citealt{Mazzali2009,Mazzali2013}), it is difficult to reconcile the data with a WD progenitor.  We therefore now examine the possibility that the progenitor of SN\,2005ek was a massive star. 

The inferred ejecta mass and abundances provide some constraints on the pre-explosion mass and envelope density structure of any massive-star progenitor for SN\,2005ek.  With typical core-collapse SN producing a compact remnant having a mass $\gtrsim 1.3$ M$_\odot$ \citep{Fryer2001}, the ejecta of SN\,2005ek would have only constituted a small fraction of the pre-explosion progenitor mass.  At the same time, the presence of a small but non-negligible amount of Ni, Si, and Mg implies that a portion of the ejecta was nuclearly processed during the explosion.  In this context, we discuss three potential massive-star explosion mechanisms for SN\,2005ek: an iron core-collapse SN, an electron-capture SN, and a fallback SN.

\subsubsection{Iron Core-Collapse SN}\label{sec:iron}
Stars with masses $\gtrsim 11$ M$_\odot$ are expected to proceed through silicon burning before undergoing an iron-core collapse at the ends of their lives.  While some amount of fallback is expected in all core-collapse SN \citep{Fryer1999,Woosley2002}, for stars with initial masses $\lesssim 20$ M$_\odot$ the final remnant mass will be dominated by the mass of the iron core at the time of collapse (i.e., approximately the Chandrasekhar mass; we will examine the case of more significant fallback in \S \ref{sec:fall}).  In this case, we would infer a pre-explosion mass for SN\,2005ek of $\sim 2$ M$_\odot$.  This may additionally apply for stars with initial masses around $\sim 50$ M$_\odot$ in the case of strong Wolf-Rayet winds (\citealt{Woosley2002}; see their Fig. 16). 

The environment of SN\,2005ek makes the latter situation (a high-mass progenitor with exceptionally strong Wolf-Rayet winds) less likely.  With a large offset from the center of its host (see \S \ref{sec:HostProps}) and the metallicity gradients common in spiral galaxies \citep{Zaritsky1994}, it is likely that SN\,2005ek exploded in a low-metallicity environment. In this case, mass loss from massive stars should be reduced rather than enhanced \citep{Vink2001}.  In addition, such a high-mass progenitor would necessarily be short lived, and the lack of H$\alpha$ flux at the explosion site is more consistent with a lower-mass progenitor ($\lesssim 12$ M$_\odot$; \S \ref{sec:HostProps}).  

For a lower-mass progenitor, it would likely be necessary to invoke binary stripping as stars of this initial mass are not expected to strip their hydrogen envelopes via winds.  This situation would be similar to that described by \citet{Nomoto1994} for SN\,1994I. In that model, a star having an initial mass of $\sim 15$ M$_\odot$ was stripped via binary interaction, yielding a C-O star of $\sim 2$ M$_\odot$ which underwent core collapse. Recall that the spectra of SN\,1994I closely resembled those of SN\,2005ek (Fig.~\ref{fig:Icspec}) and its decline rate was intermediate between SN\,2005ek and the bulk of other SN~Ib/Ic (Fig.~\ref{fig:DeltaM15}).  The ejecta mass \citet{Nomoto1994} derived for SN\,1994I was 0.88 M$_\odot$, slightly larger than that inferred for SN\,2005ek.  Taken at face value, it would be necessary to invoke either a star having a smaller initial mass (only slightly above the lower limit for iron-core collapse, consistent with the host environment described above) or more extreme stripping to reproduce SN\,2005ek.  

In either case, we would expect the amount of radioactive material produced to be on the low side for SN~Ib/Ic (due to the small amount of material at sufficient densities) and the composition to be dominated by oxygen, along with carbon and neon.  Both predictions are well matched by our observations of SN\,2005ek.  The ejecta abundances we derive for SN\,2005ek are very similar to those found for SN\,2007gr \citep{Mazzali2010} and SN\,1994I \citep{Sauer2006}.  If SN\,2005ek is due to the iron-core collapse of a massive star, it exhibits one of the lowest kinetic energies (2.5--5.2 $\times 10^{50}$ ergs) and most extreme ratios of ejecta mass to remnant mass ($M_{\rm ej}/M_{\rm remnant} < 1.0$) ever observed.  However, the inferred ratio of kinetic energy to ejecta mass is similar to that of other SN~Ic ($E_{K,51}/M_{\rm ej,\odot} \approx 1$).

Detailed modeling would be necessary to determine if the low kinetic energy inferred for SN\,2005ek is consistent with what one would expect from the iron-core collapse of a stripped low-mass progenitor.  \citet{Fryer2001} argue that although stripping the hydrogen envelope from a massive star should not significantly impact the resulting explosion energy, the same may not be true if the stripping extends into the C-O core (as is the case for SN~Ic).  In deriving the distribution of remnant masses, \citet{Fryer2001} assume an explosion energy which is proportional to the C-O core mass, an assumption which is qualitatively consistent with the low kinetic energy inferred for SN\,2005ek.

\subsubsection{Electron-Capture SN}
If the progenitor of SN\,2005ek was a relatively low-mass star, it is possible that the ensuing explosion was due to a (stripped) electron-capture SN rather than a traditional iron-core-collapse event. Stars within a narrow mass range ($\sim 8$--10 M$_\odot$; \citealt{Nomoto1987,Woosley1980,Iben1997,Kitaura2006}) are expected to undergo electron capture in their O-Ne-Mg core, decreasing pressure support, and causing the core to collapse to a NS \citep{Nomoto1987}.  \citet{Kitaura2006} have shown that the explosion proceeds in a similar manner to that of an iron-core-collapse SN (e.g., a delayed explosion driven by neutrinos), although a lower-energy explosion is produced and the density structure of the overlying material differs.  The explosion energy found in the models of \citet{Kitaura2006} is $\sim 10^{50}$ ergs, comparable to the kinetic energy inferred for SN\,2005ek.  However, the envelopes surrounding O-Ne-Mg cores are expected to be relatively diffuse.  This leads to an explosion which synthesizes only very small amounts of $^{56}$Ni ($\sim 10^{-3}$; \citealt{Kitaura2006,Bethe1985}) and ejecta which show few signs of nuclear processing. On both these points, our observations of SN\,2005ek are more consistent with a low-mass iron-core-collapse event than with an electron-capture SN.

\subsubsection{Fallback SN}\label{sec:fall}
Alternatively, it is possible that the low ejecta mass of SN\,2005ek is not caused by a low pre-explosion mass, but by a significant amount of fallback onto the proto-NS.  Qualitatively, this is expected to occur when the binding energy of the outer envelope is high, although the actual fallback criteria are complex, depending on the evolution of the shock velocity within the envelope as well as parameters such as metallicity and rotation.  It has been theorized that significant fallback should occur for stars with masses $\gtrsim 30$ M$_\odot$ \citep{Fryer2001,MacFadyen2001}. Although the successful ejecta in such explosions can possess low kinetic energies \citep{MacFadyen2001}, events of this sort lacking a hydrogen envelope are expected to be quite faint, as a majority of the radioactive $^{56}$Ni is synthesized in the inner portions of the ejecta (which fall back onto the proto-NS).  The peak luminosity of SN\,2005ek would likely require a non-negligible amount of mixing prior to fallback.  Additionally, the environment-based arguments against a very massive progenitor for SN\,2005ek (\S \ref{sec:iron}) still hold.

\subsection{Applicability of Conclusions to Other Rapidly Declining Events}

In the discussion above we found that SN\,2005ek could potentially be explained by either the edge-lit double detonation of a low mass white dwarf, the tidal disruption of a WD by a NS, or the iron core-collapse of stripped massive star, with the latter option preferred if one takes the strong spectroscopic similarity of SN\,2005ek to normal SN~Ic to be an indication of a similar progenitor channel.  However, we note that these conclusions may not broadly apply to all of the rapidly declining SN~I in the literature to date.   While SN2005ek, SN\,2010X, SN\,2002bj, SN\,1885A, and SN\,1939B all exhibit similar post-maximum decline rates and have low inferred ejecta masses ($\lesssim$ 0.3 M$_\odot$), only SN\,2010X possess a similar peak luminosity and spectroscopic evolution to SN\,2005ek.  SN\,2002bj, SN\,1939B, SN\,1885 are all significantly ($\gtrsim$ 1.3 mag) more luminous.  Coupled with their fast decline rate it is unclear whether a similar modeling scheme evoked in \S \ref{sec:explosion} would produce a self-consistent solution for these events.  SN\,2002bj also shows significantly bluer colors, slower expansion velocities, and a distinct spectroscopic evolution near maximum.  In addition while we argue that the explosion site of SN\,2005ek does not rule out the possibility of a massive star progenitor (by analogy with with explosion sites of many SN~IIP), SN\,1939B exploded in an elliptical galaxy and SN\,1885A (which exploded in the bulge of M31) shows no signs of a NS in its remnant \citep[see,][]{Perets2011}.  In short, while the conclusions above likely also apply to the rapidly-evolving SN\,2010X, their validity with respect to SN\,2002bj, SN\,1885A, and SN\,1939B is less clear.   More detailed modeling will be required to distinguish various possibilities.  Such modeling is currently underway for SN\,2010X (Kleiser et al.\ \emph{in prep.}).

\section{Summary and Conclusions}

We have presented the discovery and extensive multi-wavelength observations of the rapidly evolving Type I SN\,2005ek.  Here we summarize our main conclusions.

\begin{itemize}

\item Reaching a peak of $M_R = -17.3$ mag and declining by $\sim 3$ mag in the first 15 days post-maximum, SN\,2005ek is one of the fastest declining SN~I known thus far.  
\item Late-time photometric detections show a shallower decay timescale which is similar to the late-time evolution of other SN~I.
\item The spectra of SN\,2005ek closely resemble those of other SN~Ic in both morphology and velocity.  However, SN\,2005ek enters the optically thin phase at a much earlier epoch.  We present evidence for the onset of nebular spectroscopic features at only 9 days post-maximum.
\item The bolometric light curve of SN\,2005ek peaks at $\sim 10^{41}$ ergs s$^{-1}$. Its evolution is consistent with an explosion powered by $\sim 0.03$ M$_\odot$ of $^{56}$Ni, with incomplete gamma-ray trapping at early times.
\item We estimate the ejecta mass and kinetic energy of SN\,2005ek to be 0.3--0.7 M$_\odot$ and 2.5--5.2 $\times$ 10$^{50}$ ergs, respectively.
\item SN\,2005ek exploded in a star forming galaxy, but with a large projected offset in an area lacking strong H$\alpha$ emission.
\item The ejecta of SN\,2005ek are dominated by oxygen ($\sim 86$\%).  Other intermediate-mass elements (C, Mg, Si, S, Ca) account for $\sim 13.5$\% of the ejecta, while iron-peak elements make up only 0.5\%.  These oxygen-dominated ejecta are inconsistent with the helium-shell detonation model (``.Ia'') which has previously been invoked to explain such rapidly declining events.
\item Many of the observed properties of SN\,2005ek could be explained by either the edge-lit double detonation of a low-mass WD or the tidal disruption of a WD by a NS.  However, if we assume that the strong spectroscopic similarities between SN\,2005ek and other normal SN~Ic (with a wide range of decline timescales and inferred ejecta masses) to be an indication of a similar progenitor channel, a WD progenitor becomes very unlikely. 
\item Our observations and modeling of SN\,2005ek are also consistent with the iron-core collapse of a low-mass star (12--15 M$_\odot$), stripped by binary interaction.  In particular, the abundances derived are very similar to those found for other SN~Ic.  In this case, SN\,2005ek may possess the most extreme ratio of ejecta mass to remnant mass observed for a core-collapse SN to date. The ratio of kinetic energy to ejecta mass is similar to that of other SN~Ic.
\item The rate of such rapidly declining SN~I is at least 1--3\% of the normal SN~Ia rate.
\item Based on their strong photometric and spectroscopic similarities, our conclusions likely also apply to the rapidly evolving Type~I SN\,2010X.  However, despite their similar decline rates, there are several important observational differences between SN\,2005ek and SN\,2002bj, SN\,1939B, and SN\,1885A.  More-detailed analysis will be required to determine if these three (more luminous) objects belong to the same class of explosions.

\end{itemize}

We thank L. Bildsten, K. Shen, and T. Janka for helpful discussions.   We are grateful to the staffs at the numerous observatories where we gathered data.   We kindly acknowledge Sung Park, Katsuki Shimasaki, and Tom Matheson for assisting in the acquisition of some of the observations presented here. We acknowledge useful conversations at a Sky House workshop.

MRD is supported in part by the NSF through a Graduate Research Fellowship.  Support for this work was provided by the David and Lucile Packard Foundation Fellowship for Science and Engineering awarded to AMS. JMS is supported by an NSF Astronomy and Astrophysics Postdoctoral Fellowship under award AST-1302771. ASF acknowledges support from the US National Science Foundation (NSF) under grant SES 1056580.

This work was supported in part by the National Science Foundation under Grant No. PHYS-1066293 and the hospitality of the Aspen Center for Physics.  We also acknowledge the hospitality of the Kavli Institute for Theoretical Physics and partial support by the National Science Foundation under Grant No. NSF PHY11-25915.

The W.~M.\ Keck Observatory, which is operated as a scientific partnership among the California Institute of Technology, the University of California, and NASA; the observatory was made possible by the generous financial support of the W.~M.\ Keck Foundation.  Observations reported here were obtained at the MMT Observatory, a joint facility of the Smithsonian Institution and the University of Arizona. The Hobby-Eberly Telescope (HET), is a joint project of the University of Texas at Austin, the Pennsylvania State University, Stanford University, Ludwig-Maximilians-Universit\"{a}t M\"{u}nchen, and Georg-August-Universit\"{a}t G\"{o}ttingen. The HET is named in honor of its principal benefactors, William P. Hobby and Robert E. Eberly.  The Very Large Array is a facility of the National Science Foundation operated under cooperative agreement by Associated Universities, Inc.  PAIRITEL is operated by the Smithsonian Astrophysical Observatory (SAO) and was made possible by a grant from the Harvard University Milton Fund, the camera loan from the University of Virginia, and the continued support of the SAO and UC Berkeley. We thank M. Skrutskie for his continued support of the PAIRITEL project.

The supernova research of A.V.F.'s group at U.C. Berkeley is supported by Gary \& Cynthia Bengier, the Richard \& Rhoda Goldman Fund, the Christopher R. Redlich Fund, the TABASGO Foundation, and NSF grant AST-1211916.  KAIT and its ongoing work were made possible by donations from Sun Microsystems, Inc., the Hewlett-Packard Company, AutoScope Corporation, Lick Observatory, the NSF, the University of California, the Sylvia \& Jim Katzman Foundation, and the TABASGO Foundation. 


\clearpage

\begin{deluxetable}{lc cccc}
\tabletypesize{\tiny}
\setlength{\tabcolsep}{0.02in}
\tablecaption{P60 Photometry \label{tab:PhotomP60}}
\tablewidth{0pt}
\tablehead{
\colhead{UT Date} &
\colhead{MJD} &
\colhead{$B$ (err)} &
\colhead{$V$ (err)} &
\colhead{$R$ (err)} &
\colhead{$I$ (err)}  \\
& & \colhead{mag} &\colhead{mag} &\colhead{mag} &\colhead{mag}
}
\startdata
2005 Sep.\ 26 &53639.3   &18.25  (0.02)  &  17.83  (0.02) & 17.46 (0.01)   &  17.20 (0.02)  \\
2005 Sep.\ 27 &53640.3   &18.38  (0.03)  &  18.03  (0.03) & 17.41 (0.02)    &  17.13 (0.04) \\
2005 Sep.\ 28 &53641.3   &18.65  (0.02)  &  17.92  (0.01) & 17.60 (0.01)    &  17.18 (0.02)  \\
2005 Sep.\ 29 &53642.2   & \nodata      &  \nodata        &  17.69 (0.02)  &   \nodata \\
2005 Sep.\ 30 &53643.2   &19.10  (0.05)  &  18.24  (0.02) & 17.86 (0.01) &  17.47 (0.03)    \\
2005 Oct.\ 02 &53645.3   &19.71  (0.07)  &  18.66  (0.02) & 18.18 (0.01)  &  17.71 (0.02) \\
2005 Oct.\ 03 &53646.5   &20.07  (0.07)  &  18.93  (0.02) &  \nodata      &   \nodata  \\  
2005 Oct.\ 05 &53648.0   &  \nodata      &  19.48  (0.06) & 18.83 (0.03)  &  18.13 (0.06)  \\
2005 Oct.\ 06 &53649.2   &20.67  (0.04)  &  19.63  (0.03) & 19.03 (0.02)  &  18.26 (0.02)   \\
2005 Oct.\ 07 &53650.4   &20.90  (0.04)  &  19.86  (0.03) & 19.26 (0.02)    &  18.51 (0.02)   \\
2005 Oct.\ 08 &53651.3   &21.05  (0.04)  &  19.98  (0.04) & 19.48 (0.02)    &  18.61 (0.02)   \\
2005 Oct.\ 09 &53652.5   &\nodata      &  20.35  (0.05) & 19.75 (0.04)    &  18.74 (0.03)   \\
2005 Oct.\ 11 &53654.2   &21.74  (0.12)  &  20.60  (0.08) & 20.08 (0.05)    &  19.01 (0.05)  \\
2005 Oct.\ 12 &53655.2   &\nodata        &  20.74  (0.10) & \nodata    &  \nodata \\
2005 Oct.\ 13 &52656.2   &\nodata        &  20.88  (0.08) & 20.47 (0.08)    &  19.47 (0.06) \\
2005 Oct.\ 14 &53657.2   &\nodata        &  21.22  (0.13) & \nodata    &  \nodata 
\enddata                   
\end{deluxetable}

\begin{deluxetable}{lc ccc}
\tabletypesize{\tiny}
\setlength{\tabcolsep}{0.02in}
\tablecaption{KAIT/P200 Photometry \label{tab:Photom}}
\tablewidth{0pt}
\tablehead{
\colhead{UT Date} &
\colhead{MJD} &
\colhead{Telescope} &
\colhead{Filter} &
\colhead{mag (err)}
}
\startdata
2005 Sep.\ 18 & 53631.5 & KAIT & clear & $\lesssim$ 19 \\
2005 Sep.\ 24 & 53637.5 & KAIT & clear & 17.82 (0.08) \\  
2005 Sep.\ 25 & 53638.4 & KAIT & clear & 17.64 (0.09) \\
2005 Sep.\ 29 & 53642.5 & KAIT & clear & 17.82 (0.07) \\
2005 Oct.\ 11 &53654.9   &  P200 & $g'$ & 21.02 (0.01) \\
2005 Oct.\ 11 &53654.9   &  P200 & $r'$ & 20.28 (0.01) \\
2005 Oct.\ 11 &53654.9   &  P200 & $i'$ & 19.96 (0.01) \\
2005 Dec.\ 05 &53709.5 &  P200 &  $g'$ & $<$22.7  \\
2005 Dec.\ 05 &53709.5 &  P200 &  $r'$ &  $<$22.3 \\
2005 Dec.\ 05 &53709.5 &  P200 &  $i'$ & 21.56 (0.03)
\enddata                   
\end{deluxetable}

\begin{deluxetable}{lc cccccc}
\tabletypesize{\tiny}
\setlength{\tabcolsep}{0.02in}
\tablecaption{UVOT Photometry \label{tab:PhotomUVOT}}
\tablewidth{0pt}
\tablehead{
\colhead{UT Date} &
\colhead{MJD} &
\colhead{$uvw2$ (err)} &
\colhead{$uvm2$ (err)} &
\colhead{$uvw1$ (err)} &
\colhead{$u$ (err)} &
\colhead{$b$ (err)} &
\colhead{$v$ (err)} \\
& & \colhead{mag} &\colhead{mag} &\colhead{mag} &\colhead{mag}
&\colhead{mag} &\colhead{mag} 
}
\startdata
2005 Sep.\ 28 &53641.8  & $<$20.6 & $<$20.1 & $<$19.8 &\nodata & \nodata & 18.10 (0.17) \\
2005 Sep.\ 29 &53642.0  & $<$20.5 & $<$20.3 & $<$20.2 &\nodata & \nodata & \nodata \\
2005 Sep.\ 29 &53642.9   & $<$20.6 & \nodata & \nodata & 19.28 (0.15) & 18.97 (0.10) & 18.40 (0.11) \\
2005 Sep.\ 30 &53643.5  &  $<$20.9 & $<$20.5 & $<$20.2 & 19.10 (0.19) & 19.24 (0.18) & 18.48 (0.19) \\
2005 Oct.\ 06 &53649.3   & $<$21.1 & & $<$20.6 & \nodata & \nodata & 19.70 (0.22) \\
2005 Oct.\ 10 &53653.5   & $<$21.0 & $<$20.5& $<$20.3 & $<$19.8 & $<$ 19.8 & $<$19.6 \\
2005 Oct.\ 16 &53659.9  & $<$21.5 & $<$19.7 & \nodata & \nodata & \nodata & $<$19.9
\enddata                   
\end{deluxetable}

\clearpage

\begin{deluxetable}{lccccc}
\tabletypesize{\scriptsize}
\tablecaption{Optical Spectroscopy \label{tab:Spectra}}
\tablewidth{0pt}
\tablehead{
\colhead{UT Date} &
\colhead{MJD} &
\colhead{Phase (d)\tablenotemark{a}} &
\colhead{Target} &
\colhead{Telescope} &
\colhead{Instrument}
 }
\startdata
2005 Sep.\ 26 & 53639 & -1 & SN\,2005ek & Shane 3-m & Kast \\
2005 Sep.\ 27 & 53640 & 0  & SN\,2005ek & Tillinghast 60-in& FAST \\
2005 Sep.\ 28 & 53641 & +1 & SN\,2005ek & Tillinghast  60-in& FAST  \\
2005 Sep.\ 30 & 53643 & +3 & SN\,2005ek & Tillinghast  60-in& FAST \\
2005 Oct.\ 01 & 53644 & +4  & SN\,2005ek & Tillinghast  60-in& FAST \\
2005 Oct.\ 03 & 53646 & +6  & SN\,2005ek & Tillinghast  60-in& FAST \\
2005 Oct.\ 06 & 53649 & +9  & SN\,2005ek & Keck-II & DEIMOS \\
2005 Oct.\ 08 & 53651& +11 & SN\,2005ek & HET & LRS  \\
2011 Feb.\ 22 & 55614 & \nodata & UGC 2526 & MMT & Blue Channel \\ 
2011 Feb.\ 23 & 55615 & \nodata & UGC 2526 & MMT & Blue Channel
\enddata
\tablenotetext{a}{With respect to $R$-band maximum light.}
\end{deluxetable}

\begin{deluxetable}{lccccc}
\tabletypesize{\scriptsize}
\tablecaption{Basic Photometric Properties \label{tab:PhotomProps}}
\tablewidth{0pt}
\tablehead{
\colhead{Band} &
\colhead{$m_{\rm obs,max}$\tablenotemark{a}} &
\colhead{$M_{\rm abs,max}$\tablenotemark{b}} &
\colhead{$\Delta m_{15}$} &
\colhead{Decline Rate\tablenotemark{c}} &
\colhead{$\tau_e$\tablenotemark{d}} \\
 & \colhead{(mag)} & \colhead{(mag)} & \colhead{(mag)} &
 \colhead{(mag day$^{-1}$)} &\colhead{(day)} 
}
\startdata
$B$  & 18.25 (0.02) & -16.72 (0.15) & 3.51 (0.13) & 0.24 (0.01) & 4.7 (0.1)\\
$V$ & 17.83 (0.02) & -16.96 (0.15) & 2.79 (0.07) & 0.22 (0.01) & 6.9 (0.1)\\
$R$ & 17.41 (0.02) & -17.26 (0.15) & 2.88 (0.05) & 0.21 (0.01) & 6.3 (0.1) \\
$I$ & 17.13 (0.04) & -17.38 (0.15) & 2.13 (0.06) & 0.15 (0.01) & 8.5 (0.2)\\
$J$ & 17.10 (0.09) & -17.26 (0.17) & 1.17 (0.50) & 0.11 (0.01) & 9.6 (0.5) \\
$H$ & 16.85 (0.10) & -17.45 (0.18) & 1.62 (0.77) & 0.09 (0.01) & 12.6 (0.7) \\
$K$ & 16.49 (0.10) & -17.78 (0.18) & 2.51 (1.12) & 0.09 (0.02) & 10.2 (0.9)
\enddata
\tablenotetext{a}{Not corrected for Milky Way extinction.}
\tablenotetext{b}{Corrected for Milky Way extinction and assuming no host-galaxy extinction.}
\tablenotetext{c}{As measured by a linear fit between +2 and +18 days (from $R$-band maximum). }
\tablenotetext{d}{Time for light-curve to decline by a factor of $1/e$.}
\end{deluxetable}

\begin{deluxetable}{llc}
\tabletypesize{\scriptsize}
\tablecaption{Derived Explosion Parameters \label{tab:ExpPara}}
\tablewidth{0pt}
\tablehead{
\colhead{Parameter} &
\colhead{Unit} &
\colhead{Value} 
}
\startdata
$L_{\rm bol,peak}$ & $10^{42}$ ergs s$^{-1}$ & $1.2 \pm 0.2$\\ 
$E_{\rm rad}$\tablenotemark{a} & $10^{47}$ ergs & $8.2 \pm 0.3$  \\
$M_{\rm ej}$ &  M$_\odot$ & 0.3--0.7 \\
$E_K$ & 10$^{51}$ erg & 0.25--0.52 \\
$M_{Ni}$ & M$_\odot$ & 0.02--0.03 \\
$v_{\rm phot,max}$ & km s$^{-1}$ & $8500 \pm 500$
\enddata
\tablenotetext{a}{Emitted between $-1$ and +16 days.}
\end{deluxetable}

\end{document}